\documentclass{lmcs}
\pdfoutput=1

\usepackage{lastpage}
\lmcsdoi{19}{2}{6}
\lmcsheading{}{\pageref{LastPage}}{}{}%
{May~06,~2022}{Apr.~25,~2023}{}

\usepackage[utf8]{inputenc}
\usepackage{amsmath,
	    amsfonts,
	    amstext,
	    amssymb,
	    mathrsfs,
	    stmaryrd,
	    latexsym,
	    enumitem,
	    proof,
	    graphicx,
	    color,
	    hyperref,
	    comment}

\usepackage{mathpartir}
\usepackage{mathtools}
\usepackage[all]{xy}
\usepackage{tikz}

\usepackage{macro}

\newcommand{\cws}[2]
	{\\ \centerline{$#2$} \\[-#1pt]}

\newcommand{\fullbox}
	{{\mbox{}\nolinebreak\hfill{$\rule{2mm}{2mm}$}}}

\newcommand{\bibtrick}[1]
	{}

\newcommand{\lap}
	{\mbox{$<$}}
\newcommand{\rap}
	{\mbox{$>$}}

\newcommand{\lsp}
	{\llbracket}
\newcommand{\rsp}
	{\rrbracket}

\newcommand{\lmp}
	{\{ \! | \,}
\newcommand{\rmp}
	{\, | \! \}}

\newcommand{\bfpi}
        {\mbox{\boldmath $\pi$}}

\newcommand{\cala}
        {\mathcal{A}}

\newcommand{\calc}
        {\mathcal{C}}

\newcommand{\calk}
        {\mathcal{K}}

\newcommand{\calm}
        {\mathcal{M}}

\newcommand{\calr}
        {\mathcal{R}}

\newcommand{\cals}
        {\mathcal{S}}

\newcommand{\calu}
        {\mathcal{U}}

\newcommand{\natns}
	{\mathbb{N}}

\newcommand{\realns}
	{\mathbb{R}}

\newcommand{\procs}
	{\mathbb{P}}

\newcommand{\auxarrow}
	{\mathop{\longrightarrow}}

\begin{document}

\title[Bridging Causal Reversibility and Time Reversibility]
      {Bridging Causal Reversibility and Time Reversibility: \\
       A Stochastic Process Algebraic Approach}

\author[M.~Bernardo]{Marco Bernardo}
\author[C.A.~Mezzina]{Claudio A.\ Mezzina}

\address{Dipartimento di Scienze Pure e Applicate, Universit\`a di Urbino, Italy}
\email{marco.bernardo@uniurb.it, claudio.mezzina@uniurb.it}

\begin{abstract}
Causal reversibility blends reversibility and causality for concurrent systems. It indicates that an action
can be undone provided that all of its consequences have been undone already, thus making it possible to
bring the system back to a past consistent state. \linebreak Time reversibility is instead considered in the
field of stochastic processes, mostly for efficient analysis purposes. A performance model based on a
continuous-time Markov chain is time reversible if its stochastic behavior remains the same when the
direction of time is reversed. We bridge these two theories of reversibility by showing the conditions under
which causal reversibility and time reversibility are both ensured by construction. This is done in the
setting of a stochastic process calculus, which is then equipped with a variant of stochastic bisimilarity
accounting for both forward and backward directions.
\end{abstract}

\keywords{Reversibility, Causality, Markov Chains, Process Calculi, Bisimilarity}

\maketitle

%
%
\section{Introduction}
\label{sec:intro}
%
%

The interest into computation reversibility dates back to the 60’s, when it was observed that irreversible
computations cause heat dissipation into circuits~\cite{Landauer61}. More precisely, Landauer's principle
states that any logically irreversible manipulation of information, such as the erasure of bits or the
merging of computation paths, must be accompanied by a corresponding entropy increase in
non-information-bearing degrees of freedom of the information processing apparatus or its
environment~\cite{Bennett03}. Hence, according to this principle, which has been recently verified
in~\cite{BerutAPCDL12} and given a physical foundation in~\cite{Frank18}, any logically reversible
computation, in which no information is erased, may be potentially carried out without releasing any heat.
This suggested that low energy consumption could be achieved by resorting to \emph{reversible computing}, in
which there is no information loss~\cite{Bennett73}. Nowadays, reversible computing has several applications
such as biochemical reaction modeling~\cite{PhillipsUY12,Pinna17}, parallel discrete-event
simulation~\cite{PerumallaP14,SchordanOJB18}, robotics~\cite{LaursenES18}, control theory~\cite{SiljakPP19},
fault tolerant systems~\cite{DanosK05,deVriesKH10,LaneseLMSS13,VassorS18}, and concurrent program
debugging~\cite{GiachinoLM14,LaneseNPV18a}.

In a reversible system, we see two directions of computation: a \emph{forward} one, coinciding \linebreak
with the normal way of computing, and a \emph{backward} one, which is able to undo the effects of the
forward one when needed. In the literature, there are different variants of reversibility. For instance, in
a Petri net reversibility means that one can always reach the initial marking~\cite{BarylskaKMP18}, while in
a distributed system it stands for the capability of returning to a past consistent
state~\cite{rccs,DanosK05}. In contrast, in the performance evaluation field reversibility is related to
time and is instrumental to develop efficient analysis methods~\cite{Kelly79}.

Our focus is on integrating \emph{causal reversibility} and \emph{time reversibility}. Causal reversibility
describes the capability of going back to a past state that is consistent with the computational history of
a system; in this setting, quantitative aspects have been totally disregarded. On the other hand, the theory
of time reversibility studies the conditions under which the stochastic behavior of a system remains the
same when the direction of time is reversed; unfortunately, it has been applied to concurrent systems
without explicitly taking causality into account. In this paper, we aim at bridging these two theories by
showing how causal reversibility and time reversibility can be jointly achieved. To this purpose, since
process algebra constitutes a common ground for concurrency theory and probability
theory~\cite{LarsenS91,Hillston96}, we develop our proposal by considering a stochastic process calculus, in
which actions are equipped with positive real numbers. Each of these numbers expresses the rate at which
\linebreak the corresponding action is executed and uniquely identifies the exponential distribution that
quantifies the duration of the action, so that the stochastic process underlying such a calculus turns out
to be a continuous-time Markov chain (CTMC)~\cite{KemenyS60}.

	\begin{figure}[t]

\centerline{\includegraphics{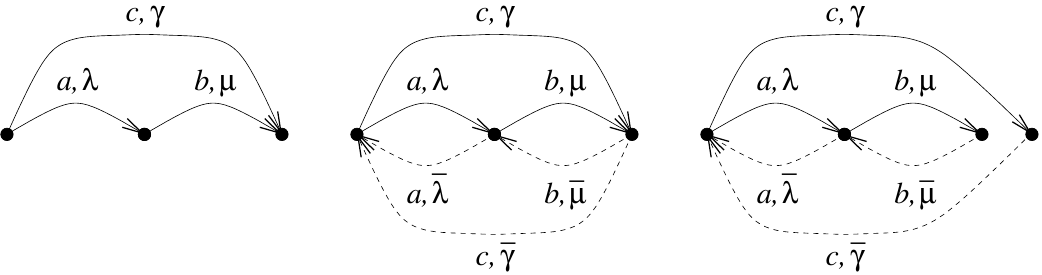}}
\caption{Making a system causally reversible and time reversible}
\label{fig:causal_time_rev}

        \end{figure}

Ensuring that a system is both causally reversible and time reversible is not a trivial task. \linebreak
Consider for instance a system that can perform either action $a$ at rate $\lambda$ followed by action~$b$
at rate $\mu$, or action $c$ at rate $\gamma$. In stochastic process algebra terms, it would be represented
as $\lap a, \lambda \rap . \lap b, \mu \rap . \nil + \lap c, \gamma \rap . \nil$ where ``.'' is the action
prefix operator, ``+'' is the choice operator, and $\nil$ is the terminated process. The underlying labeled
transition system is the action-labeled CTMC depicted in the leftmost part of
Figure~\ref{fig:causal_time_rev}. This system is not reversible because, given a pair of states connected by
a transition, it is not possible to go from the target state back to the source state. So the first step
towards reversibility is to add a backward transition for each such pair of states, which is labeled with
the same action as the forward transition together with a suitable backward rate. The resulting labeled
transition system is the action-labeled CTMC depicted in the central part of
Figure~\ref{fig:causal_time_rev}.

However, this system is not causally reversible because after performing $a$ and $b$ (resp.~$c$) the system
could go back by undoing $c$ (resp.\ $b$ and $a$), i.e., the actions in the backward direction could be
different from those in the forward direction. In order for this system to be time reversible, the product
of the rates along any cycle should be the same when changing direction~\cite{Kelly79}. This holds for
cycles of length $2$ or $4$ involving only actions $a$ and~$b$, whereas it is not the case in general with
cycles of length $3$ involving all the actions, unless $\lambda \cdot \mu \cdot \bar{\gamma} = \gamma \cdot
\bar{\mu} \cdot \bar{\lambda}$ as happens, e.g., when $\lambda = \bar{\lambda}$, $\mu = \bar{\mu}$, and
$\gamma = \bar{\gamma}$, i.e., when every backward rate is equal to the corresponding forward rate. The
version depicted in the rightmost part of Figure~\ref{fig:causal_time_rev}, which is obtained from $\lap a,
\lambda \rap . \lap b, \mu \rap . \nil + \lap c, \gamma \rap . \nil$ \linebreak by applying the approach we
propose, is both causally reversible and time reversible.

The contribution of this paper is threefold. Firstly, we apply for the first time the general methodology
for reversing process calculi of~\cite{ccsk} to a stochastic process calculus. In particular, we provide
forward and backward operational semantic rules -- featuring forward and backward actions and rates -- and
then we show that the resulting calculus is causally reversible. The latter is accomplished by importing
from the reduction semantics setting of~\cite{rccs} the notion of concurrent transitions, which is new in
the structural operational semantics framework of~\cite{ccsk} and is then handled through the recent
technique of~\cite{LanesePU20}.

Secondly, we prove that time reversibility can be obtained by using in the operational semantic rules
backward rates equal to the corresponding forward rates or by restricting the syntax in such a way that
parallel composition cannot occur within the scope of action prefix or choice. This is quite different from
the approach adopted for example in~\cite{Harrison03,MarinR15}, where time reversibility is verified a
posteriori, as we instead produce a calculus in which time reversibility can be guaranteed by construction.
The difference in the approach does not prevent us from importing in our setting compositionality results
from those works.

Thirdly, we address behavioral equivalences for our reversible stochastic process calculus, in order to
provide a means to identify systems possessing the same functional and performance properties. In
particular, we focus on Markovian bisimilarity~\cite{Hillston96}, which equates systems capable of stepwise
mimicking each other's functional and performance behavior. We extend a result of~\cite{DeNicolaMV90} to our
stochastic setting by showing that a variant of Markovian bisimilarity on computations accounting for both
forward and backward directions coincides with Markovian bisimilarity on states, thus inheriting the nice
properties of the latter~\cite{Ber07}.

This paper, which is a revised and enriched version of~\cite{BernardoM20}, is organized as follows. In
Sections~\ref{sec:causal_rev} and~\ref{sec:time_rev} we recall background notions about causal reversibility
and time reversibility, respectively. In Section~\ref{sec:rmpc_integration} we develop our proposal of
integration for these two forms of reversibility in the setting of a CTMC-based process calculus, which is
then equipped in Section~\ref{sec:bisim} with a forward and backward Markovian bisimilarity. In
Section~\ref{sec:examples} we provide some examples. Finally, in Section~\ref{sec:concl} we conclude with
some directions for future work.

%
%
\section{Causal Reversibility of Concurrent Systems}
\label{sec:causal_rev}
%
%

Reversibility in a computing system has to do with the possibility of reverting the last performed action.
In a sequential system, this is very simple as there exists just one last action, hence the only challenge
is how to store the information needed to reverse this last action. After Bennett~\cite{Bennett73}, several
techniques have been developed to reverse the computations of \linebreak a sequential
program~\cite{Leeman86,YokoyamaG07,HaySchmidtGCH21,PalaciosV15}.

In a concurrent system, the situation is much more complex as the last action may not be uniquely
identifiable. Indeed, there might be several concurrent last actions. One could resort to timestamps to
decide which action is the last one, but having synchronized clocks in a distributed system is rather
difficult. A good approximation is to consider as last action \linebreak every action that has not caused
any other action yet. This is at the basis of the so called \emph{causal reversibility}, which relates
reversibility with causality~\cite{rccs,DanosK05}. Intuitively, the definition states that, in a concurrent
system, any executed action can be undone provided that all of its consequences, if any, have been undone
beforehand. 

In the process algebra literature, specifically for CCS~\cite{Milner89}, two approaches have been developed
to reverse a computation and to keep track of past actions: the \emph{dynamic} approach of~\cite{rccs} and
the \emph{static} approach of~\cite{ccsk}. Despite these two approaches being quite different, they have
been recently shown to be equivalent in terms of labeled transition system isomorphism~\cite{LaneseMM21}. On
the application side, the former approach is more suitable for systems whose operational semantics is given
in terms of reduction semantics, hence it is to be preferred in the case of very expressive
calculi~\cite{LaneseMS10,CristescuKV13,LaneseM20} as well as programming
languages~\cite{LienhardtLMS12,LaneseNPV18b}. On the other hand, the latter approach is very handy when it
comes to deal with labeled transition systems and basic process calculi.

The dynamic approach relies on RCCS~\cite{rccs}, a variant of CCS that uses external memories attached to
processes to record all the actions executed by the processes themselves. Every RCCS process is thus of the
form $m \triangleright P$, where $m$ is a memory whilst $P$ is a CCS process. At the beginning of the
computation, if $P$ is not sequential, i.e., it contains occurrences of the parallel composition operator,
then an empty memory is distributed among all the sequential components. Every time that one of the
sequential components performs an action, the action is pushed on top of the memory stack of the component
as it is its last executed action. This construction is shown to allow a concurrent system to backtrack
along any causally equivalent path in such a way that no previously inaccessible state is reached, which is
the essence of causal reversibility.

In contrast, the static approach of~\cite{ccsk} proposes a general method, of which CCSK is a result, to
reverse process calculi whose operational semantic rules are expressed in the path format
of~\cite{BaetenV93}. The basic idea of this method is to make all the operators of the calculus static and
each executed action univocally identified by a communication key. In particular, since dynamic process
algebraic operators like action prefix ``.'' and choice ``+'' would disappear after a transition -- thus
causing information loss -- all of them are kept within target processes of transitions. This is the
approach that we will follow for defining our reversible stochastic process calculus in
Section~\ref{sec:rmpc_integration}.

For example, consider the CCS process $a . P + b . Q$, which either performs action $a$ and then behaves as
process $P$, or performs action $b$ and then behaves as process $Q$. In the CCS transition $a . P + b . Q \:
\xrightarrow{a} \: P$, both the executed action $a$ and the discarded process $b . Q$ are lost after
performing the transition, hence it is not possible to get back to the initial state. In CCSK the same
process evolves instead as $a . P + b . Q \: \xrightarrow{a\decor{[i]}} \: \decor{a[i] .} P \decor{\, + \, b
. Q}$, where $\decor{a[i] .} P \decor{\, + \, b . Q}$ behaves like $P$ in further forward transitions while
the colored parts can be seen as decorations of $P$ to be used only in backward transitions. Alternatively,
$\decor{a[i] .} P \decor{\, + \, b . Q}$ can be seen as $\calc[P]$, where $P$ is the active part and $\calc
= a[i] . \bullet + \, b . Q$ is its dead context. In this way the use of external memories of~\cite{rccs} is
avoided, because all the necessary information for enabling reversibility is syntactically maintained within
processes. Note that action~$a$ is decorated with a fresh key $\decor{[i]}$ -- e.g., a natural number -- so
as to distinguish among several actions executed in the past that have the same name. As we will see in
Section~\ref{sec:rmpc_integration}, this is important for correctly managing communications when going
backward.

%
%
\section{Time Reversibility of Markov Chains}
\label{sec:time_rev}
%
%

In the performance evaluation field, a different notion of reversibility, called \emph{time reversibility},
is considered. We illustrate it in the specific case of continuous-time Markov chains, which are
discrete-state stochastic processes characterized by the \emph{memoryless property}~\cite{KemenyS60}.

A \emph{stochastic process} describes the evolution of some random phenomenon over time through a set of
random variables, one for each time instant. A stochastic process $X(t)$ taking values from a discrete state
space $\cals$ for $t \in \realns_{\ge 0}$ is a \emph{continuous-time Markov chain (CTMC)} iff for all $n \in
\natns$, time instants $t_{0} < t_{1} < \dots < t_{n} < t_{n + 1} \in \realns_{\ge 0}$, and states $s_{0},
s_{1}, \dots, s_{n}, s_{n + 1} \in \cals$ it holds that $\Pr \{ X(t_{n + 1}) = s_{n + 1} \mid X(t_{i}) =
s_{i}, 0 \le i \le n \} = \Pr \{ X(t_{n + 1}) = s_{n + 1} \mid X(t_{n}) = s_{n} \}$, i.e., the probability
of moving from one state to another does not depend on the particular path that has been followed in the
past to reach the current state, hence that path can be forgotten.

A CTMC can be equivalently represented as a labeled transition system or as a state-indexed matrix. In the
first case, each transition is labeled with some probabilistic information describing the evolution from the
source state $s$ to the target state $s'$ of the transition itself. \linebreak In the second case, the same
information is stored into an entry, indexed by those two states, of a square matrix. The value of this
information is, in general, a function of time.

We restrict ourselves to \emph{time-homogeneous} CTMCs, in which conditional probabilities of the form $\Pr
\{ X(t + t') = s' \mid X(t) = s \}$ do not depend on $t$, so that the information considered above is given
by $\lim_{t' \rightarrow 0} \frac{\Pr \{ X(t + t') = s' \mid X(t) = s \}}{t'}$. This limit yields a number
called the \emph{rate} at which the CTMC moves from state $s$ to state~$s'$ and characterizes the
exponentially distributed random time taken by the considered move. It can be shown that the sojourn time in
any state $s \in \cals$ is exponentially distributed with rate given by the sum of the rates of the moves
of~$s$. The average sojourn time in $s$ is the inverse of such a sum and the probability of moving from $s$
to $s'$ is the ratio of the corresponding rate to the aforementioned sum.

A CTMC is \emph{irreducible} iff each of its states is reachable from every other state with probability
greater than $0$. A state $s \in \cals$ is \emph{recurrent} iff the CTMC will eventually return to~$s$ with
probability $1$, in which case $s$ is called \emph{positive recurrent} iff the expected number of steps
until the CTMC returns to it is finite. A CTMC is \emph{ergodic} iff it is irreducible and all of its states
are positive recurrent; ergodicity coincides with irreducibility in the case that the CTMC has finitely many
states, as they form a finite strongly connected component.

Every time-homogeneous and ergodic CTMC $X(t)$ is \emph{stationary}, which means that $(X(t_{i} + t'))_{1
\le i \le n}$ has the same joint distribution as $(X(t_{i}))_{1 \le i \le n}$ for all $n \in \natns_{\ge 1}$
and $t_{1} < \dots < t_{n}, t' \in \realns_{\ge 0}$. In this case, $X(t)$ has a unique \emph{steady-state
probability distribution} $\bfpi = (\pi(s))_{s \in \cals}$ that fulfills $\pi(s) = \lim_{t \rightarrow
\infty} \Pr \{ X(t) = s \mid X(0) = s' \}$ for any $s' \in \cals$ because the CTMC has reached equilibrium.
These probabilities are computed by solving the linear system of \emph{global balance equations} $\bfpi
\cdot \mathbf{Q} = \mathbf{0}$ subject to $\sum_{s \in \cals} \pi(s) = 1$ and $\pi(s) \in \realns_{> 0}$
\linebreak for all $s \in \cals$. The \emph{infinitesimal generator matrix} $\mathbf{Q} = (q_{s, s'})_{s, s'
\in \cals}$ contains for each pair of distinct states the rate of the corresponding move, which is $0$ in
the absence of a direct move between them, while $q_{s, s} = - \sum_{s' \neq s} q_{s, s'}$ for all $s \in
\cals$, i.e., every diagonal element contains the opposite of the total exit rate of the corresponding
state, so that each row of $\mathbf{Q}$ sums up to $0$. Therefore $\bfpi \cdot \mathbf{Q} = \mathbf{0}$
means that, once reached equilibrium, for every state the incoming probability flux is equal to the outgoing
probability flux.

Due to state space explosion and numerical stability problems, the calculation of the solution of the global
balance equation system is not always feasible~\cite{Stewart94}. However, it can be tackled in the case that
the behavior of the considered CTMC remains the same when the direction of time is reversed. A CTMC $X(t)$
is \emph{time reversible} iff $(X(t_{i}))_{1 \le i \le n}$ has the same joint distribution as $(X(t' -
t_{i}))_{1 \le i \le n}$ for all $n \in \natns_{\ge 1}$ and $t_{1} < \dots < t_{n}, t' \in \realns_{\ge 0}$.
\linebreak In this case $X(t)$ and its reversed version $X^{\textrm{r}}(t) = X(-t)$, $t \in \realns_{\ge
0}$, are stochastically identical, in particular they are stationary and share the same steady-state
probability distribution $\bfpi$. \linebreak In order for a stationary CTMC $X(t)$ to be time reversible, it
is necessary and sufficient that the \emph{partial balance equations} $\pi(s) \cdot q_{s, s'} \, = \,
\pi(s') \cdot q_{s', s}$ are satisfied for all $s, s' \in \cals$ such that $s \neq s'$ or, equivalently,
that $q_{s_{1}, s_{2}} \cdot \ldots \cdot q_{s_{n - 1}, s_{n}} \cdot q_{s_{n}, s_{1}} \, = \, q_{s_{1},
s_{n}} \cdot q_{s_{n}, s_{n - 1}} \cdot \ldots \cdot q_{s_{2}, s_{1}}$ \linebreak for all $n \in \natns_{\ge
2}$ and distinct $s_{1}, \dots, s_{n} \in \cals$~\cite{Kelly79}. Note that the sum of the partial balance
equations for $s \in \cals$ yields the global balance equation $\pi(s) \cdot |q_{s, s}| \, = \, \sum_{s'
\neq s} \pi(s') \cdot q_{s', s}$.

The time-reversed version $X^{\textrm{r}}(t)$ of a stationary CTMC $X(t)$ can be defined even when $X(t)$ is
not reversible. As shown in~\cite{Kelly79,Harrison03}, this is accomplished by using the steady-state
probability distribution $\bfpi$ of $X(t)$, with $X^{\textrm{r}}(t)$ turning out to be a CTMC too and having
the same steady-state probability distribution $\bfpi$. More precisely, $q_{s_{j}, s_{i}}^{\textrm{r}} =
\frac{\pi(s_{i})}{\pi(s_{j})} \cdot q_{s_{i}, s_{j}}$ for all $s_{i} \neq s_{j}$, i.e., the rate from state
$s_{j}$ to state $s_{i}$ in the time-reversed CTMC is proportional to the rate from state $s_{i}$ to state
$s_{j}$ in the original CTMC, where the coefficient is given by the ratio of $\pi(s_{i})$ to $\pi(s_{j})$.
Note that the time-reversed version of $X^{\textrm{r}}(t)$ is $X(t)$.

%
%
\section{Integrating Causal Reversibility and Time Reversibility}
\label{sec:rmpc_integration}
%
%

In this section, we integrate the two concepts of causal reversibility and time reversibility recalled in
the previous two sections by means of a simple calculus named RMPC -- Reversible Markovian Process Calculus.
We start with its syntax -- where actions are paired with rates -- and its semantics, which are inspired
by~\cite{ccsk}. We then show that the reversibility induced by RMPC is consistent with causality by adapting
the notion of concurrent transitions from~\cite{rccs} and exploiting the technique of~\cite{LanesePU20}.
Finally, we exhibit the conditions under which time reversibility is achieved too and we compare our
setting, in which time reversibility is ensured by construction, with those of~\cite{Harrison03,MarinR15},
in which time reversibility is verified a posteriori, and import from them some results enabling an
efficient analysis of performance.

%
\subsection{Syntax of RMPC}
\label{sec:integr_calculus_syn}
%

The syntax of RMPC is shown in Table~\ref{tab:syntax}. A standard \emph{forward process} $P$ can be one of
the following: the terminated process $\nil$; the action-prefixed process $\lap a, \lambda \rap . P$, which
is able to perform an action $a$ at rate $\lambda$ -- called an exponentially timed action as its duration
follows an exponential distribution of parameter~$\lambda$ -- and then continues as process $P$; the choice
$P + Q$ between processes $P$ and $Q$ based on the rates of their initial actions; or the parallel
composition $P \coop{L} Q$, indicating that processes $P$ and $Q$ execute in parallel and must synchronize
only on actions belonging to $L$.

A \emph{reversible process} $R$ is built on top of forward processes. As in~\cite{ccsk}, the syntax of
reversible processes differs from the one of forward processes due to the fact that in the former each
prefix $\lap a, \lambda \rap$ can be decorated with a \emph{communication key} $i$ thus becoming $\lap a,
\lambda \rap[i]$. A process of the form $\lap a, \lambda \rap[i] . R$ expresses that in the past the process
synchronized with the environment and this synchronization was identified by key $i$. Keys are thus attached
only to already executed actions; they are not needed in the absence of parallel composition, provided that
executed actions are marked in some way.

	\begin{table}[t]

\[\begin{array}{|rcl|}
\hline
P, Q & \!\! \sdef \!\! & \nil \mid \lap a, \lambda \rap . P \mid P + Q \mid P \coop{L} Q \\[0.1cm]
R, S & \!\! \sdef \!\! & P \mid \lap a, \lambda \rap[i] . R \mid R + S \mid R \coop{L} S \, \\
\hline
\end{array}\]

\caption{Syntax of forward processes (top) and reversible processes (bottom)}
\label{tab:syntax}

	\end{table}

Let $\Act$ be a countable set of actions (ranged over by $a, b$), $\Rate = \realns_{> 0}$ be a set of rates
(ranged over by $\lambda, \mu$), and $\Key$ be a countable set of keys (ranged over by $i, j$). Let $\Lbl =
\Act \times \Rate \times \Key$ be a set of labels each formed by an action, a rate, and a communication key
(ranged over by $\ell$). Given a forward label $\ell = \lap a, \lambda \rap[i]$, we write $\past{\ell} =
\lap a, \past{\lambda} \rap[i]$ for the corresponding backward label, where $\past{\lambda}$ stands for the
backward rate of the action -- i.e., the rate at which the action is undone -- whose value will be discussed
later on.

We denote by $\Procs$ the set of processes generated by the production for $R$ in Table~\ref{tab:syntax},
while we use predicate $\std(\_)$ to identify the standard forward processes that can be \linebreak derived
from the production for $P$ in the same table. For example, $\lap a,\lambda \rap.\lap b,\mu \rap.\nil$ is a
standard forward process that can perform action $a$ at rate $\lambda$ followed by action $b$ at rate $\mu$,
\linebreak while $\lap a,\lambda \rap[i].\lap b,\mu \rap.\nil$ is a non-standard reversible process that can
either undo action $a$ at rate $\past{\lambda}$ or perform action $b$ at rate $\mu$. Note that $\lap
a,\lambda \rap.\lap b,\mu \rap[j].\nil \, \notin \, \Procs$ because a future action cannot precede a past
action in the description of the behavior of a process.

The set $\key(R)$ of keys occurring in a process $R \in \Procs$ is inductively defined as follows:
\cws{6}{\begin{array}{rcl}
\key(P) & \!\! = \!\! & \emptyset \\
\key(\lap a,\lambda \rap[i].R) & \!\! = \!\! & \{i\} \cup \key(R) \\
\key(R + S) & \!\! = \!\! & \key(R) \cup \key(S) \\
\key(R \coop{L} S) & \!\! = \!\! & \key(R) \cup \key(S) \\
\end{array}}

%
\subsection{Semantics for RMPC}
\label{sec:integr_calculus_sem}
%

The semantics for RMPC is the labeled transition system $(\Procs, \Lbl, \reduction{\phantom{aa}})$. As
in~\cite{ccsk}, the transition relation $\reduction{\phantom{aa}} \; \subseteq \Procs \times \Lbl \times
\Procs$ is given by $\reduction{\phantom{aa}} \; = \; \fw{\;\;\;\;} \cup \bk{\;\;\;} \!$, \linebreak where
the \emph{forward transition relation} $\fw{\;\;\;}$ and the \emph{backward transition relation}
$\bk{\;\;\;} \!\!$ are \linebreak the least relations respectively induced by the forward rules in the left
part of Table~\ref{tab:sem} and by the backward rules in the right part of the same table.

	\begin{table}[t]

\[\begin{array}{|ll|}
\hline
\inferrule*[left=Act1]{\std(R)}{\lap a,\lambda \rap.R \fw{<a,\lambda> [i]} \lap a,\lambda \rap[i].R}
&
\inferrule*[left=Act1$^\rev$]{\std(R)}{\lap a,\lambda \rap[i].R \bk{<a,\past{\lambda}>[i]} \lap a,\lambda
\rap.R} \\[0.3cm]
\inferrule*[left=Act2]{R \fw{<b,\mu>[j]} R' \hspace{0.25cm} j \neq i}{\lap a,\lambda \rap[i].R
\fw{<b,\mu>[j]} \lap a,\lambda \rap[i].R'}
&
\inferrule*[left=Act2$^\rev$]{R \bk{<b,\past{\mu}>[j]} R' \hspace{0.25cm} j \neq i}{\lap a,\lambda \rap[i].R
\bk{<b,\past{\mu}>[j]} \lap a,\lambda \rap[i].R'} \\[0.3cm]
\inferrule*[left=Cho]{R \fw{<a,\lambda>[i]} R' \hspace{0.25cm} \std(S)}{R+S \fw{<a,\lambda>[i]} R'+S}
&
\inferrule*[left=Cho$^\rev$]{R \bk{<a,\past{\lambda}>[i]} R' \hspace{0.25cm} \std(S)}{R+S
\bk{<a,\past{\lambda}>[i]} R'+S} \\[0.3cm]
\inferrule*[left=Par]{R \fw{<a,\lambda>[i]} R' \hspace{0.25cm} a \notin L \hspace{0.25cm} i \notin
\key(S)}{R \coop{L} S \fw{<a,\lambda>[i]} R' \coop{L} S}
&
\inferrule*[left=Par$^\rev$]{R \bk{<a,\past{\lambda}>[i]} R' \hspace{0.25cm} a \notin L \hspace{0.25cm} i
\notin \key(S)}{R \coop{L} S \bk{<a,\past{\lambda}>[i]} R' \coop{L} S} \\[0.3cm]
\inferrule*[left=Coo]{R \fw{<a,\lambda>[i]} R' \hspace{0.25cm} S \fw{<a,\mu>[i]} S' \hspace{0.25cm} a \in
L}{R \coop{L} S \fw{<a,\lambda\cdot\mu>[i]} R' \coop{L} S'}
&
\inferrule*[left=Coo$^\rev$]{R \bk{<a,\past{\lambda}>[i]} R' \hspace{0.25cm} S \bk{<a,\past{\mu}>[i]} S'
\hspace{0.25cm} a \in L}{R \coop{L} S \bk{<a,\past{\lambda}\cdot\past{\mu}>[i]} R' \coop{L} S'} \\
\hline
\end{array}\]

\caption{Structural operational semantic rules for RMPC}
\label{tab:sem}

	\end{table}

Rule \textsc{Act1} handles processes of the form $\lap a, \lambda \rap.P$, where $P$ is written as $R$
subject to $\std(R)$. In addition to transforming the action prefix into a transition label, it generates a
key $i$ that is bound to the action $\lap a, \lambda \rap$ thus yielding the label $\lap a, \lambda \rap
[i]$. As can be noted, the prefix is not discarded by the application of the rule, instead it becomes a
key-storing part of the target process that is necessary to offer again that action after coming back. Rule
\textsc{Act1}$^\rev$ reverts the action $\lap a, \lambda \rap [i]$ of the process $\lap a, \lambda \rap [i]
. R$ provided that $R$ is a standard process, which ensures that $\lap a, \lambda \rap [i]$ is the only past
action that is left to undo. One of the main design choices of the entire framework is how the rate
$\past{\lambda}$ of the backward action is calculated. For the time being, we leave it unspecified in
\textsc{Act1}$^\rev$ as the value of this rate is not needed to prove the causal consistency part of
reversibility; we will see later on \linebreak that it may be important to achieve time reversibility.

The presence of rules \textsc{Act2} and \textsc{Act2}$^\rev$ is motivated by the fact that rule
\textsc{Act1} does not discard the executed prefix from the process it generates. In particular, rule
\textsc{Act2} allows a process $\lap a,\lambda \rap[i].R$ to execute if $R$ itself can execute provided that
the action performed by~$R$ picks a key $j$ different from~$i$, so that all the action prefixes in a
sequence are decorated with distinct keys. Rule \textsc{Act2}$^\rev$ simply propagates the execution of
backward actions from inner subprocesses that are not standard as long as key uniqueness is preserved, in
such a way that past actions are overall undone from the most recent one to the least recent one.

Unlike the classical rules for the choice operator~\cite{Milner89}, rule \textsc{Cho} does not discard the
part of the overall process that has not contributed to the executed action. More in detail, if process $R$
does an action, say $\lap a,\lambda \rap[i]$, and becomes $R'$, then the entire process $R + S$ becomes $R'
+ S$ as the information about $+ \, S$ is necessary for offering again the original choice after coming
back. Once the choice is made, only the non-standard process $R'$ can proceed further, with the standard
process $S$ constituting a dead context of $R'$. Note that, in order to apply rule \textsc{Cho}, at least
one of $R$ and $S$ has to be standard, meaning that it is impossible for two non-standard processes to
execute if they are composed by a choice operator. Rule \textsc{Cho}$^\rev$ has precisely the same structure
as rule \textsc{Cho}, but deals with the backward transition relation; if $R'$ is standard, then the dead
context $S$ will come into play again. For both rules, we omit their symmetric variants in which it is $S$
to move.

The semantics of parallel composition is inspired by~\cite{Hoare85}. Rule \textsc{Par} allows process~$R$
within $R \coop{L} S$ to individually perform an action $\lap a, \lambda \rap [i]$ provided that $a \notin
L$. It is also checked that the executing action is bound to a fresh key $i \notin \key(S)$, thus ensuring
the uniqueness of communication keys across parallel composition too. Rule \textsc{Coo} instead allows both
$R$ and $S$ to move by synchronizing on any action in the set $L$, provided that the communication key is
the same on both sides, i.e., $i \in \key(R') \cap \key(S')$. The resulting cooperation action has the same
name and the same key and is assumed to be exponentially distributed with rate given by the product of the
rates of the two involved actions~\cite{Hillston94}. Rules \textsc{Par}$^\rev$ and \textsc{Coo}$^\rev$
respectively have the same structure as \textsc{Par} and \textsc{Coo}; the symmetric variants of
\textsc{Par} and \textsc{Par}$^\rev$ are omitted.

In our stochastic setting, choice and parallel composition are not subject to nondeterminism. The decision
of which action is chosen or which process advances is made probabilistically based on the rates of the
actions that are ready to execute. This mechanism is called the \emph{race policy}~\cite{Hillston96} because
the higher the rate, i.e., the faster the action, the higher its execution probability. This is consistent
with the fact that the minimum of a set of exponentially distributed random variables is still exponentially
distributed, from which it follows that, as already mentioned in Section~\ref{sec:time_rev}, the sojourn
time in any state of a CTMC is exponentially distributed with rate given by the sum of the rates of the
outgoing transitions. The average sojourn time in that state is the inverse of such a sum and the
probability of moving from that state to another one is the ratio of the rate of the corresponding
transition to the aforementioned sum.

We further recall from~\cite{Hillston96} that the race policy and the memoryless property of CTMCs allow the
\emph{interleaving view of concurrency} -- typically employed in nondeterministic process calculi -- to be
adopted in the setting of a stochastic process calculus like RMPC. Consider the standard process $\lap a,
\lambda \rap . \nil \coop{\emptyset} \lap b, \mu \rap . \nil$. It can initially execute either action~$a$ at
rate $\lambda$, which happens with probability $\frac{\lambda}{\lambda + \mu}$, or action $b$ at rate $\mu$,
which happens with probability $\frac{\mu}{\lambda + \mu}$; the average sojourn time in the corresponding
state is $\frac{1}{\lambda + \mu}$. Observing that the probability of simultaneous termination of the two
actions is zero, suppose that $a$ terminates first. In the state reached by the corresponding forward
transition labeled with $a$ and $\lambda$ the execution of $b$ is still in progress, so there will be an
outgoing forward transition labeled with $b$ and a parameter quantifying the residual duration of $b$.  That
parameter is still $\mu$ because, thanks to the memoryless property, the residual time to the termination of
$b$ is still exponentially distributed with rate $\mu$ (see the left part of the forthcoming
Figure~\ref{fig:expansion_law}).

%
\subsection{Reachable Processes}
\label{sec:integr_calculus_reach_proc}
%

The syntax in Table~\ref{tab:syntax} prevents future actions from preceding past actions. However, this is
not the only necessary limitation, because not all the processes generated by the considered grammar are
semantically meaningful. On the one hand, in the case of a choice at least one of the two subprocesses must
be standard, hence for instance $\lap a, \lambda \rap [i] . \nil + \lap b, \mu \rap [j] . \nil$ is not
admissible as it indicates that both branches have been selected. On the other hand, key uniqueness must be
enforced within non-standard processes, so for example $\lap a, \lambda \rap [i] . \lap b, \mu \rap [i] .
\nil$ and $\lap a, \lambda \rap [i] . \nil \coop{\emptyset} \lap b, \mu \rap [i] . \nil$ are not admissible
either.

In the rest of the paper, we thus consider only \emph{reachable processes}, whose set we denote by $\procs$.
They include processes from which a computation can start, i.e., standard forward processes, as well as
processes that can be derived from the previous ones via finitely many applications of the rules for
$\fw{\;\;\;\;}$ in~Table~\ref{tab:sem}. Given a reachable process $R \in \procs$, we observe that if
$\std(R)$ then $\key(R) = \emptyset$ while any other process $R'$ reachable from $R$ is such that $\key(R')
\neq \emptyset$, as at least one of the actions occurring in $R$ has been executed and hence \linebreak it
has been equipped with a communication key inside $R'$.

%
\subsection{Preliminary Reversibility Properties}
\label{sec:integr_prelim}
%

A basic property in order for RMPC to be reversible, both in the causal sense and in the time one, is the so
called \emph{loop lemma}~\cite{rccs,ccsk}. This property states that each executed action can be undone and
that any undone action can be redone. In other words, when considering the states associated with two
arbitrary reachable processes, either there is no transition between them, or there is a pair of transitions
such that one is a forward transition from the first state to the second state while the other is a backward
transition from the second state to the first state.

	\begin{lem}[\emph{loop lemma}]\label{lm:loop}
Let $R, S \in \procs$. Then $R \fw{<a,\lambda>[i]} S$ iff $S \bk{<a,\past{\lambda}>[i]} R$.

		\begin{proof}
By induction on the depth of the derivation of the considered transition by noting that each forward (resp.\
backward) rule has a corresponding backward (resp.\ forward) rule.
		\end{proof}

	\end{lem}

Given a sequence $\sigma = \ell_1 \dots \ell_n$ of $n \in \natns_{> 0}$ labels, $R \fw{\sigma} S$ denotes a
sequence of forward transitions $R \fw{\ell_1} R_1 \fw{\ell_2} \cdots \fw{\ell_n} S$ labeled with that
sequence. If $\past{\sigma} = \past{\ell_n} \dots \past{\ell_1}$ is the corresponding backward sequence,
then for each $\ell_i$ occurring in $\sigma$ it holds that $R_{i-1} \fw{\ell_i} R_i$ iff $R_i
\bk{\past{\ell_i}} R_{i-1}$. The loop lemma thus generalizes as follows.
 
	\begin{cor}\label{cor:loop}
Let $R, S \in \procs$. Then $R \fw{\sigma} S$ iff $S \bk{\past{\sigma}} R$.
\fullbox

	\end{cor}

%
\subsection{Causal Reversibility of RMPC}
\label{sec:integr_causal_rev}
%

We now prove the causal reversibility of RMPC, which means that each of its reachable processes is able to
backtrack \emph{correctly}, i.e., without encountering previously inaccessible states, and \emph{flexibly},
i.e., along any causally equivalent path. \linebreak To this purpose, we start by borrowing some machinery
from~\cite{rccs}, in particular the notion of concurrent transitions, that needs to be adapted because the
reversing method of~\cite{ccsk} \linebreak we have followed is different from the one of~\cite{rccs} we are
going to exploit.

Given a transition $\theta = R \reduction{\ell} S$ with $R, S \in \procs$, we call $R$ the \emph{source}
of~$\theta$ and $S$ its \emph{target}. If $\theta$ is a forward transition, i.e., $\theta = R \fw{\ell} S$,
we denote by $\past{\theta} = S \bk{\, \past{\ell}} R$ the corresponding backward transition. Two
transitions are said to be \emph{coinitial} if they have the same source and \emph{cofinal} if they have the
same target. Two transitions are \emph{composable} when the target of the first transition coincides with
the source of the second transition. A finite sequence of pairwise composable transitions is called a
\emph{computation}. We use $\et$ for the empty computation and $\trace$ to range over computations, with
$|\trace|$ denoting the length of $\trace$ expressed as the number of transitions constituting it. When
$\trace$ is a forward computation, we denote by $\past{\trace}$ the corresponding backward computation. The
notions of source, target, coinitiality, cofinality, and composability naturally extend to computations. We
indicate with $\trace_1 \trace_2$ the composition of the two computations $\trace_1$ and $\trace_2$ when
they are composable.

Before specifying when two transitions are concurrent, we need to present the set of causes -- identified by
keys -- that lead to a given communication key, along with the notion of process context.

The \emph{causal set} $\cau(R,i)$ of process $R \in \procs$ until key $i \in \calk$ is inductively defined
as:
\cws{0}{\begin{array}{rcll}
\cau(P, i) & \!\! = \!\! & \emptyset \\
\cau(\lap a, \lambda \rap [j] . R, i) & \!\! = \!\! & \left\{ \begin{array}{ll}
\emptyset & \textrm{if } j = i \textrm{ or } i \notin \key(R) \\
\{j\} \cup \cau(R, i) & \textrm{otherwise} \\
\end{array}
\right. \\
\cau(R + S, i) & \!\! = \!\! & \cau(R, i) \cup \cau(S, i) & \\
\cau(R \coop{L} S, i) & \!\! = \!\! & \cau(R, i) \cup \cau(S, i) & \\
\end{array}}
If $i \in \key(R)$, then $\cau(R, i)$ represents the set of keys in $R$ that caused $i$, with $\cau(R, i)
\subset \key(R)$ because on the one hand $i \notin \cau(R, i)$ and on the other hand keys that are not
causally related to $i$ are not considered. A key $j$ causes $i$ if it appears syntactically before $i$ in
$R$ or, said otherwise, $i$ is inside the scope of $j$. 

A \emph{process context} is a process with a hole in it, generated by the grammar:
\cws{0}{\calc \: \sdef \: \bullet \mid \lap a, \lambda \rap [i] . \calc \mid R + \calc \mid \calc + R \mid R
\coop{L} \calc \mid \calc \coop{L} R}
We write $\calc[R]$ to denote the process obtained from $\calc$ by replacing its hole with $R$.

We are now in a position to define what we mean by concurrent transitions on the basis of the notion of
conflicting transitions. The first condition below tells that a forward transition is in conflict with a
backward one whenever the latter tries to undo a cause of the key of the former. The second condition below
deems as conflictual two transitions respectively generated by the two subprocesses of a choice.

	\begin{defi}[\emph{conflicting and concurrent transitions}]\label{def:conf}
Two coinitial transitions $\theta_1$ and $\theta_2$ from a process $R \in \procs$ are in conflict iff one of
the following two conditions holds:

		\begin{enumerate}

\item $\theta_1 = R \fw{<a, \lambda>[i]} S_1$ and $\theta_2 = R \bk{<b, \past{\mu}>[j]} S_2$ with $j \in
\cau(S_1, i)$.

\item $R = \calc[P_1 + P_2]$ with $\theta_k$ deriving from $P_k \fw{<a_k, \lambda_k>[i_k]} S_k$ for $k = 1,
2$.

		\end{enumerate}

\noindent
Two coinitial transitions are concurrent when they are not in conflict.
\fullbox

	\end{defi}

	\begin{figure}

\[
\xymatrix{
	&\lap b,\mu \rap.\lap a,\lambda \rap.P\\
	\lap b,\mu \rap[j].\lap a,\lambda \rap.P \ar@{-->}^{<b,\past{\mu}>[j]} [ur]
	\ar@{->}^{<a,\lambda>[i]}[r] & 
	\lap b,\mu \rap[j].\lap a,\lambda \rap[i].P\\
	&\lap a,\lambda \rap.P + \lap a,\lambda \rap.P \ar@{->}_{<a,\lambda>[i]} [dl]
	\ar@{->}^{<a,\lambda>[i]} [dr]\\
	\lap a,\lambda \rap[i].P+\lap a,\lambda \rap.P && \lap a,\lambda \rap.P+ \lap a,\lambda \rap [i].P
	}
\]

\caption{Two examples of conflicting transitions}
\label{fig:conf_tr}

	\end{figure}

Figure~\ref{fig:conf_tr} shows two related examples. In the former, the process $\lap b, \mu \rap [j] . \lap
a, \lambda \rap . P$ can perform two transitions: a backward one and a forward one. They are in conflict
according to the first condition of Definition~\ref{def:conf} as the backward transition removes the key $j$
that is in the causal set of $i$. In the latter, we have that process $\lap a, \lambda \rap . P + \lap a,
\lambda \rap . P$ is able to trigger two forward transitions. Since they arise from the same choice
operator, they are in conflict according to the second condition of Definition~\ref{def:conf}.

	\begin{rem}
In a stochastic process calculus like RMPC the semantic treatment of the choice operator is not trivial
because for a process of the form $\lap a, \lambda \rap . P + \lap a, \lambda \rap . P$ \linebreak the
operational rules should produce either a single $a$-transition whose rate is $\lambda + \lambda$, or two
$a$-transitions each having rate $\lambda$ that do not collapse into a single one~\cite{Hillston96}. In our
reversible framework, two distinct transitions are naturally generated thanks to the fact that the key
associated with the executed action and the discarded alternative subprocess are stored into the reached
process, as shown in the bottom part of Figure~\ref{fig:conf_tr}.
\fullbox

	\end{rem}
 
We finally show that reversibility is causally consistent in our concurrent framework. This can be done in
two ways: either by adapting the original proof of~\cite{rccs}, as we did in~\cite{BernardoM20}, or by using
the general technique recently provided by~\cite{LanesePU20}. We opt for the latter, according to which
causal consistency stems from the \emph{diamond property} -- which amounts to concurrent transitions being
confluent -- \emph{backward transition independence} -- which generalizes the concept of backward
determinism used for reversible sequential languages~\cite{YokoyamaG07} -- and \emph{past well foundedness}
-- which ensures that reachable processes have a finite past.

Concurrent transitions can commute as formally stated below by the diamond property (as an example see the
left part of the forthcoming Figure~\ref{fig:expansion_law}), while conflicting ones cannot.

	\begin{lem}[\emph{diamond property}]\label{lm:diamond}
Let $\theta_1 = R \reduction{\ell_1} S_1$ and $\theta_2 = R \reduction{\ell_2} S_2$ be two coinitial
transitions from a process $R \in \procs$. If $\theta_1$ and $\theta_2$ are concurrent, then there exist two
cofinal transitions $\theta'_2 = S_1 \reduction{\ell_2} S$ and $\theta'_1 = S_2 \reduction{\ell_1} S$ with
$S \in \procs$.

		\begin{proof}
The proof is by case analysis on the direction of $\theta_1$ and $\theta_2$. We distinguish three cases
according to whether the two transitions are both forward, both backward, or one forward and the other
backward. Suppose that $\ell_1 = \lap a,\lambda \rap[i]$ and $\ell_2 = \lap b,\mu \rap[j]$, with $i \neq j$
otherwise $\theta_1$ and $\theta_2$ would be generated by the two subprocesses of a choice operator and
hence could not be concurrent:

			\begin{itemize}

\item Suppose that $\theta_1$ and $\theta_2$ are both forward. Since $\theta_1$ and $\theta_2$ are
concurrent, by virtue of Definition~\ref{def:conf} the two transitions cannot originate from a choice
operator. They must thus be generated by a parallel composition, but not through the \textsc{Coo} rule
because $\theta_1$ and $\theta_2$ have different keys and hence cannot synchronize. Without loss of
generality, \linebreak we can assume that $R = R_1 \coop{L} R_2$ with $R_1 \fw{<a,\lambda>[i]} S_1$ and $R_2
\fw{<b,\mu>[j]} S_2$ and $a,b \not\in L$. \linebreak By applying the \textsc{Par} rule, we have that $R_1
\coop{L} R_2 \fw{<a,\lambda>[i]} S_1 \parallel_{L} R_2 \fw{<b,\mu>[j]} S_1 \parallel_{L} S_2$ \linebreak as
well as $R_1 \coop{L} R_2 \fw{<b,\mu>[j]} R_1 \parallel_{L} S_2 \fw{<a,\lambda>[i]} S_1 \parallel_{L} S_2$.

\item The case in which $\theta_1$ and $\theta_2$ are both backward is similar to the previous one.

\item Suppose that $\theta_1$ is forward and $\theta_2$ is backward. Since $\theta_1$ and $\theta_2$ are
concurrent, by virtue of Definition~\ref{def:conf} the backward transition cannot remove a cause of the
forward one. Since either subprocess of a choice operator or a parallel composition cannot perform a forward
transition and a backward transition without preventing the backward one from removing a cause of the
forward one, and in the case of the choice operator only one of the two subprocesses can perform
transitions, without loss of generality we can assume that $R = R_1 \coop{L} R_2$ with $R_1
\fw{<a,\lambda>[i]} S_1$ and $R_2 \bk{<b,\past{\mu}>[j]} S_2$ as well as $a,b \not\in L$ so as to preserve
causes. By applying the \textsc{Par} rule, we have that $R_1 \coop{L} R_2 \fw{<a,\lambda>[i]} S_1
\parallel_{L} R_2 \linebreak \bk{<b,\past{\mu}>[j]} S_1 \parallel_{L} S_2$ as well as $R_1 \coop{L} R_2
\bk{<b,\past{\mu}>[j]} R_1 \parallel_{L} S_2 \fw{<a,\lambda>[i]} S_1 \parallel_{L} S_2$.
\qedhere

			\end{itemize}

		\end{proof}

	\end{lem} 

	\begin{lem}[\emph{backward transition independence}]
Let $R \in \procs$. Then any two coinitial backward transitions $\theta_1 = R \bk{<a,\past{\lambda}>[i]}
S_1$ and $\theta_2 = R \bk{<b,\past{\mu}>[j]} S_2$ are concurrent.

		\begin{proof}
Since by Definition~\ref{def:conf} there is no case in which two backward transitions are conflicting, the
property trivially holds.
		\end{proof}

	\end{lem}

	\begin{lem}[\emph{past well foundedness}]
Let $R_0 \in \procs$. Then there is no infinite sequence of backward transitions such that $R_i \bk{\ell_i}
R_{i+1}$ for all $i \in \natns$.

		\begin{proof}
It can be easily proved by induction on $|\key(R_0)|$ by observing that a backward transition decreases by
one the total number of keys of $R_0$, with this number being finite.
		\end{proof}

	\end{lem}

Following~\cite{rccs,Levy76}, we also define a notion of \emph{causal equivalence} over computations, which
abstracts from the order of concurrent transitions. In this way, computations obtained by swapping the order
of their concurrent transitions are identified with each other (see for instance the left part of the
forthcoming Figure~\ref{fig:expansion_law}) and the composition of a computation with its inverse is
identified with the empty computation.

	\begin{defi}[\emph{causal equivalence}]
Causal equivalence is the smallest equivalence relation $\ceq$ on computations that is closed under
composition and satisfies the following:
		\begin{enumerate}

\item $\theta_1 \theta'_2 \ceq \theta_2 \theta'_1$ for any two coinitial concurrent transitions $\theta_1 =
R \reduction{\ell_1} R_1$ and $\theta_2 = R \reduction{\ell_2} R_2$ and any two cofinal transitions
$\theta'_2 = R_1 \reduction{\ell_2} S$ and $\theta'_1 = R_2 \reduction{\ell_1} S$ respectively composable
with the previous ones.

\item $\theta \past{\theta} \ceq \et$ and $\past{\theta} \theta \ceq \et$.
\fullbox

		\end{enumerate}

	\end{defi}


The further property below, called \emph{parabolic lemma} in~\cite{LanesePU20}, states that any computation
can be seen as a backward computation followed by a forward computation. As observed in~\cite{rccs}, up to
causal equivalence one can always reach for the maximum freedom of choice among transitions by going
backward and only then going forward (not the other way around). Intuitively, one could depict computations
as parabolas: the system first draws potential energy from its memory, by undoing all the executed actions,
and then restarts.

	\begin{lem}[\emph{parabolic lemma}]
For any computation $\omega$, there exist two forward computations $\omega_1$ and $\omega_2$ such that
$\omega \ceq \past{\omega_1} \omega_2$ and $|\omega_1| + |\omega_2| \leq |\omega|$.

		\begin{proof}
It follows from the diamond property and backward transition independence thanks to~\cite{LanesePU20}.
		\end{proof}

	\end{lem}

	\begin{thm}[\emph{causal consistency}]\label{thm:causal_cons_rev}
Let $\trace_1$ and $\trace_2$ be two computations. Then $\trace_1 \ceq \trace_2$ iff $\trace_1$ and
$\trace_2$ are both coinitial and cofinal.

		\begin{proof}
It follows from past well foundedness and the parabolic lemma thanks to~\cite{LanesePU20}.
		\end{proof}

	\end{thm}

Theorem~\ref{thm:causal_cons_rev} shows that causal equivalence characterizes a space for admissible
rollbacks that are (i)~correct as they do not lead to states not reachable by some forward computation and
(ii)~flexible enough to allow undo operations to be rearranged with respect to the order in which the undone
concurrent transitions were originally performed. This implies that the states reached by any backward
computation could be reached by performing forward computations only. Therefore, we can conclude that RMPC
meets causal reversibility.

%
\subsection{Time Reversibility of RMPC}
\label{sec:integr_time_rev}
%

The rules in Table~\ref{tab:sem} associate with any process $R \in \procs$ a labeled transition system $\lsp
R \rsp = (\procs, \Lbl, \reduction{\phantom{aa}}, R)$ whose initial state corresponds to~$R$. To investigate
time reversibility, we have to derive from $\lsp R \rsp$ the underlying CTMC $\calm \lsp R \rsp$ and we have
to specify the value of any rate $\past{\lambda}$ labeling a backward transition.

First of all, we observe that every state of $\lsp R \rsp$ with an outgoing forward transition actually has
infinitely many copies of that transition. The motivation is that rules \textsc{Act1} and \textsc{Act2}
generate a transition for each possible admissible key, with the key being part of both the label and the
reached process. For example, $\lap a, \lambda \rap . P$ has an outgoing forward transition towards $\lap a,
\lambda \rap [i] . P$ for each $i \in \calk$; note that from each $\lap a, \lambda \rap [i] . P$ there is
instead a single outgoing backward transition to $\lap a, \lambda \rap . P$. When building the CTMC
underlying $\lap a, \lambda \rap . P$, only one of those outgoing forward transitions has to be considered,
otherwise the corresponding state would have an infinite exit rate.

In the construction of $\calm \lsp R \rsp$ we thus need to reason in terms of transition bundles. \linebreak
A \emph{transition bundle} collects all the transitions departing from the same state and labeled with the
same action, the same rate, and different keys, whose target states are syntactically identical up to keys
in the same positions. To this purpose, we make use of contexts to introduce $\equiv_{\calk}$ as the
smallest equivalence relation on~$\procs$ that satisfies the following:

	\begin{itemize}

\item If $\calc$ does not contain occurrences of $\coop{L}$ with $a \in L$ and $i$ and $j$ do not occur in
$\calc$ and $R$, then $\calc[\lap a, \lambda \rap [i] . R] \: \equiv_{\calk} \: \calc[\lap a, \lambda \rap
[j] . R]$.

\item If $a \in L_{1} \cap \ldots \cap L_{n - 1}$ for $n \ge 2$, $i$ and $j$ do not occur in $\calc$,
$\calc_{1}$, $\dots$, $\calc_{n}$, $R_{1}$, $\dots$, $R_{n}$, and no \linebreak further $a$ with key $i$
occurs in $\calc$, then $\calc[\calc_{1}[\lap a, \lambda_{1} \rap [i] . R_{1}] \coop{L_{1}} \dots \coop{L_{n
- 1}} \calc_{n}[\lap a, \lambda_{n} \rap [i] . R_{n}]] \linebreak \equiv_{\calk} \: \calc[\calc_{1}[\lap a,
\lambda_{1} \rap [j] . R_{1}] \coop{L_{1}} \dots \coop{L_{n - 1}} \calc_{n}[\lap a, \lambda_{n} \rap [j] .
R_{n}]]$.

	\end{itemize}

\noindent
As for the second case, for instance $\lap a, \lambda_{1} \rap . P_{1} \coop{\{ a \}} \lap a, \lambda_{2}
\rap . P_{2}$ evolves to $\lap a, \lambda_{1} \rap [i] . P_{1} \coop{\{ a \}} \lap a, \lambda_{2} \rap [i] .
P_{2}$ for all $i \in \calk$, but we cannot replace $\lap a, \lambda_{1} \rap [i] . P_{1}$ with $\lap a,
\lambda_{1} \rap [j] . P_{1}$, unless we replace $\lap a, \lambda_{2} \rap [i] . P_{2}$ with $\lap a,
\lambda_{2} \rap [j] . P_{2}$ too, because $\lap a, \lambda_{1} \rap [j] . P_{1} \coop{\{ a \}} \lap a,
\lambda_{2} \rap [i] . P_{2}$ has no backward transition to $\lap a, \lambda_{1} \rap . P_{1} \coop{\{ a \}}
\lap a, \lambda_{2} \rap . P_{2}$ when $j \neq i$.

Among the representations mentioned in Section~\ref{sec:time_rev}, below we choose to formalize $\calm \lsp
R \rsp$ as a labeled transition system because this is closer to $\lsp R \rsp$ obtained from the rules in
Table~\ref{tab:sem}.

	\begin{defi}[\emph{underlying CTMC}]\label{def:underl_ctmc}
The CTMC underlying a process $R \in \procs$ is the labeled transition system $\calm \lsp R \rsp = (\procs /
\! \equiv_{\calk}, \cala \times \calr, \reduction{\phantom{aa}}_{\calk}, [R]_{\equiv_{\calk}})$ where:

		\begin{itemize}

\item $\procs / \! \equiv_{\calk}$ is the quotient set of $\equiv_{\calk}$ over $\procs$, i.e., the set of
classes of $\equiv_{\calk}$-equivalent processes, representing the set of states.

\item $[R]_{\equiv_{\calk}}$ is the $\equiv_{\calk}$-equivalence class of $R$, which simply is $\{ R \}$
when $R$ is standard and hence contains no keys, representing the initial state.

\item $\reduction{\phantom{aa}}_{\calk} \; \subseteq (\procs / \! \equiv_{\calk}) \times (\cala \times
\calr) \times (\procs / \! \equiv_{\calk})$ is the transition relation given by
$\reduction{\phantom{aa}}_{\calk} \; = \; \fw{\;\;\;\;}_{\calk} \cup \bk{\;\;\;}_{\!\! \calk}$:

			\begin{itemize}

\item $[S]_{\equiv_{\calk}} \fw{<a, \lambda>}_{\calk} [S']_{\equiv_{\calk}}$ whenever $S \fw{<a,
\lambda>[i]} S'$ for some $i \in \calk$.

\item $[S]_{\equiv_{\calk}} \bk{<a, \past{\lambda}>}_{\!\! \calk \,\,} [S']_{\equiv_{\calk}}$ whenever $S
\bk{<a, \past{\lambda}>[i]} S'$ for some $i \in \calk$.
\fullbox

			\end{itemize}

		\end{itemize}

	\end{defi}

When moving from $\lsp R \rsp$ to $\calm \lsp R \rsp$, individual states are thus replaced by classes of
states that are syntactically identical up to keys in the same positions; moreover, keys are removed from
transition labels. As a consequence, every state of $\calm \lsp R \rsp$ turns out to have finitely many
outgoing transitions. We also note that $\calm \lsp R \rsp$ is an action-labeled CTMC, as each of its
transitions is labeled not only with a rate but also with an action.

As a preliminary step towards time reversibility, we have to show that $\calm \lsp R \rsp$ is stationary. It
holds that $\calm \lsp R \rsp$ is even ergodic thanks to the loop lemma.

	\begin{lem}\label{lm:ergodicity}
Let $R \in \procs$. Then $\calm \lsp R \rsp$ is time homogeneous and ergodic.

		\begin{proof}
The time homogeneity of $\calm \lsp R \rsp$ is a straightforward consequence of the fact that its rates
simply are positive real numbers, not time-dependent functions. The ergodicity of $\calm \lsp R \rsp$ stems
from the fact that the graph representing $\calm \lsp R \rsp$ is a finite strongly connected component due
to Corollary~\ref{cor:loop}.
		\end{proof}

	\end{lem}

The proof of time reversibility exploits the necessary and sufficient condition based on partial balance
equations~\cite{Kelly79} (see Section~\ref{sec:time_rev}). The loop lemma and the assumption $\past{\lambda}
= \lambda$ for all $\lambda \in \realns_{> 0}$ ensure that the steady-state probability distribution of
$\calm \lsp R \rsp$ is the uniform distribution. From this it immediately follows that the partial balance
equations are satisfied, i.e., that time reversibility holds.

	\begin{lem}\label{lm:unif_stat_distr}
Let $R \in \procs$, $\cals$ be the set of states of $\calm \lsp R \rsp$, and $n = |\cals|$. If every
backward rate is equal to the corresponding forward rate, then the steady-state probability distribution
$\bfpi$ of $\calm \lsp R \rsp$ satisfies $\pi(s) = 1 / n$ for all $s \in \cals$.

		\begin{proof}
If $n = 1$, i.e., $R$ is equal to $\nil$ or to the parallel composition of several standard processes whose
initial actions have to synchronize but are different from each other, then trivially $\pi(s) = 1 / n = 1$
for the only state $s \in \cals$. \\
Suppose now that $n \ge 2$. From Lemma~\ref{lm:ergodicity} it follows that $\calm \lsp R \rsp$ has a unique
steady-state probability distribution $\bfpi$. Due to Lemma~\ref{lm:loop}, each outgoing forward/backward
transition of an arbitrary state $s \in \cals$ has a corresponding incoming backward/forward transition,
hence the global balance equation for $s$ reformulated in terms of transitions looks as follows, where $s'$
(resp., $s''$) is the target state of a forward (resp., backward) transition departing from $s$:
\cws{0}{\pi(s) \cdot (\sum\limits_{s \fw{<a', \lambda'>}_{\calk} s'} \lambda' + \sum\limits_{s \bk{<a'',
\past{\lambda''}>}_{\,\, \calk} s''} \past{\lambda''}) \: = \: \sum\limits_{s' \bk{<a',
\past{\lambda'}>}_{\,\, \calk} s} \pi(s') \cdot \past{\lambda'} + \sum\limits_{s'' \fw{<a'',
\lambda''>}_{\calk} s} \pi(s'') \cdot \lambda''}
Since every backward rate is equal to the corresponding forward rate, the global balance equation for $s$
boils down to:
\cws{0}{\pi(s) \cdot (\sum\limits_{s \fw{<a', \lambda'>}_{\calk} s'} \lambda' + \sum\limits_{s \bk{<a'',
\lambda''>}_{\,\, \calk} s''} \lambda'') \: = \: \sum\limits_{s' \bk{<a', \lambda'>}_{\,\, \calk} s} \pi(s')
\cdot \lambda' + \sum\limits_{s'' \fw{<a'', \lambda''>}_{\calk} s} \pi(s'') \cdot \lambda''}
Since the two summations related to $s'$ as well as the two summations related to $s''$ have the same number
of summands, the equation above is satified when $\pi(s) = \pi(s') = \pi(s'')$ \linebreak for each $s'$
reached by a forward transition of $s$ and each $s''$ reached by a backward transition of $s$. All global
balance equations are thus satisfied when $\pi(s) = 1 / n$ for all $s \in \cals$.
		\end{proof}

	\end{lem}

	\begin{thm}[\emph{time reversibility 1}]\label{thm:time_rev_1}
Let $R \in \procs$. If every backward rate is equal to the corresponding forward rate, then $\calm \lsp R
\rsp$ is time reversible.

		\begin{proof}
Let $\cals$ be the set of states of $\calm \lsp R \rsp$ and $n = |\cals|$; to avoid trivial cases, suppose
$n \ge 2$. From Lemma~\ref{lm:ergodicity} it follows that $\calm \lsp R \rsp$ has a unique steady-state
probability distribution~$\bfpi$. Now consider $s, s' \in \cals$ with $s \neq s'$ connected by transitions.
The proof resembles the one of Lemma~\ref{lm:unif_stat_distr}, but focuses on partial balance equations
rather than on global ones. \\
Due to Lemma~\ref{lm:loop}, the partial balance equation for $s$ and $s'$ reformulated in terms of
transitions looks as follows:
\cws{0}{\pi(s) \cdot (\sum\limits_{s \fw{<a', \lambda'>}_{\calk} s'} \lambda' + \sum\limits_{s \bk{<a'',
\past{\lambda''}>}_{\,\, \calk} s'} \past{\lambda''}) \: = \: \pi(s') \cdot (\sum\limits_{s' \bk{<a',
\past{\lambda'}>}_{\,\, \calk} s} \past{\lambda'} + \sum\limits_{s' \fw{<a'', \lambda''>}_{\calk} s}
\lambda'')}
Since every backward rate is equal to the corresponding forward rate, the partial balance equation for $s$
and~$s'$ boils down to:
\cws{0}{\pi(s) \cdot (\sum\limits_{s \fw{<a', \lambda'>}_{\calk} s'} \lambda' + \sum\limits_{s \bk{<a'',
\lambda''>}_{\,\, \calk} s'} \lambda'') \: = \: \pi(s') \cdot (\sum\limits_{s' \bk{<a', \lambda'>}_{\,\,
\calk} s} \lambda' + \sum\limits_{s' \fw{<a'', \lambda''>}_{\calk} s} \lambda'')}
Since $\pi(s) = \pi(s') = 1 / n$ due to Lemma~\ref{lm:unif_stat_distr} and the two left summations on both
sides as well as the two right summations on both sides have the same number of summands, the equation above
is satified. The result then follows from the fact that $s$ and $s'$ are two arbitrary distinct states
connected by transitions.
		\end{proof}

	\end{thm}

The result above holds under the assumption $\past{\lambda} = \lambda$ for all $\lambda \in \realns_{> 0}$.
This constraint on rates can be relaxed if we exploit the structure of the graph representing $\calm \lsp R
\rsp$. For example, it is well known that CTMCs in the form of stationary birth-death processes are time
reversible~\cite{Kelly79}, where a birth-death process comprises a totally ordered set of states such that
every state different from the final one has a (birth) transition to the next state and every state
different from the initial one has a (death) transition to the previous state. Time reversibility extends to
tree-like birth-death processes~\cite{Kelly79}, where each such variant comprises a partially ordered set of
states such that every non-final state may have several birth transitions and every non-initial state has
one death transition to its only parent state.

	\begin{lem}\label{lm:birth_death_tree}
Let $R \in \procs$. If parallel composition does not occur in $R$, then $\calm \lsp R \rsp$ is a tree-like
birth-death process.

		\begin{proof}
The fragment of graph representing $\calm \lsp R \rsp$ that includes only forward transitions, i.e.,
generated by applying only the rules in the left part of Table~\ref{tab:sem}, is a tree in the absence of
parallel composition inside $R$. In fact, observing that if $R$ is $\nil$ then the graph contains a single
state with no transitions, in any state other than $\nil$ there are two options:

			\begin{itemize}

\item The application of rule \textsc{Act1} or \textsc{Act2} yields a new transition towards a new state due
to the generation of a fresh communication key that is stored within the new state.

\item The application of rule \textsc{Cho} yields a single new transition or two new transitions --
depending on whether a single subprocess is standard or both subprocesses are standard -- each of which
reaches a new state due to the generation of a fresh communication key that is stored within the new state
and the fact that the dead context -- i.e., the standard subprocess that stays idle -- is stored within the
new state too (this is important in order for the two new states to be different when the two subprocesses
are standard and identical, see the bottom part of Figure~\ref{fig:conf_tr}).

			\end{itemize}

For each forward transition generated by one of the three forward rules mentioned above, the corresponding
backward rule generates the corresponding backward transition. Therefore $\calm \lsp R \rsp$ is a tree-like
birth-death process.
		\end{proof}

	\end{lem}

	\begin{thm}[\emph{time reversibility 2}]\label{thm:time_rev_2}
Let $R \in \procs$. If parallel composition does not occur in~$R$, then $\calm \lsp R \rsp$ is time
reversible.

		\begin{proof}
A straightforward consequence of Lemma~\ref{lm:birth_death_tree} above and Lemma~$1.5$ of~\cite{Kelly79}
establishing the time reversibility of any stationary tree-like birth-death process. Note that unlike $\calm
\lsp R \rsp$ the graph mentioned in the latter lemma is not directed, i.e., each of its edges can be
traversed in both directions, but this is the same as having two directed edges in opposite directions like
a forward transition and the corresponding backward transition.
		\end{proof}

	\end{thm}

The time reversibility of CTMC-based compositional models of concurrent systems has already been
investigated in~\cite{Harrison03}. That work examines conditions relying on the conservation of total exit
rates of states in addition to the conservation of products of rates around cycles~\cite{Kelly79} (see
Section~\ref{sec:time_rev}), which support the hierarchical and compositional reversal of stochastic process
algebra terms. These conditions also lead to the efficient calculation of steady-state probability
distributions in a product form typical of queueing theory~\cite{Kleinrock75}, thus avoiding the need of
solving the global balance equations of the entire system. More recently, in~\cite{MarinR15} similar
conditions have been employed to characterize the class of $\rho$-reversible stochastic automata, which
allow for state permutations. Under certain constraints, the joint steady-state probability distribution of
the composition of two such automata is the product of the steady-state probability distributions of the two
automata.

The main difference between our approach to time reversibility and the aforementioned ones  is twofold.
Firstly, our approach is part of a more general framework in which also causal reversibility is addressed.
Secondly, our approach is inspired by the idea of~\cite{ccsk} of developing a formalism in which it is
possible to express models whose reversibility is guaranteed by construction, instead of building a
posteriori the time-reversed version of a certain model like in~\cite{Harrison03} or verifying a posteriori
whether a given model is time reversible or not like in~\cite{MarinR15}.

It is worth noting that these methodological differences do not prevent us from importing in our setting
some results from~\cite{Harrison03,MarinR15}, although a few preliminary observations about notational
differences are necessary. Both~\cite{Harrison03} and~\cite{MarinR15} make a distinction between active
actions, each of which is given a rate, and passive actions, each of which is given a weight, with the
constraint that, in case of synchronization, the rate of the active action is multiplied by the weight of
the corresponding passive action. In RMPC there is no such distinction, however the same operation, i.e.,
multiplication, is applied to the rates of two synchronizing actions. A passive action can thus be rendered
as an action with rate $1$, while a set of alternative passive actions can be rendered as a set of actions
whose rates sum up to $1$. Moreover, in~\cite{MarinR15} synchronization is enforced between any
active-passive pair of identical actions, whereas in RMPC the parallel composition operator is enriched with
an explicit synchronization set $L$, which yields as a special case the aforementioned synchronization
discipline when $L$ is equal to the set $\Act$ of all the actions. We can therefore conclude that our
parallel composition operator is a generalization of those used in~\cite{Harrison03,MarinR15}, hence the
recalled notational differences do not hamper the transferral of results.

In~\cite{Harrison03} the compositionality of a CTMC-based stochastic process calculus is exploited to prove
RCAT -- Reversed Compound Agent Theorem, which establishes the conditions under which the time-reversed
version of the cooperation of two processes is equal to the cooperation of the time-reversed versions of
those two processes. The application of RCAT leads to product-form solution results from stochastic process
algebraic models, including a new different proof of Jackson's theorem for product-form queueing
networks~\cite{Jackson63}.

In~\cite{MarinR15} the notion of $\rho$-reversibility is introduced for stochastic automata, which are
essentially action-labeled CTMCs. Function $\rho$ is a state permutation that ensures (i)~for each action
the equality of the total exit rate of any state~$s$ and $\rho(s)$ and (ii)~the conservation of
action-related rate products across cycles when considering states in the forward direction and their
$\rho$-counterparts in the backward direction. For any ergodic $\rho$-reversible automaton, it turns out
that $\pi(s) = \pi(\rho(s))$ for every state $s$. Moreover, the synchronization inspired by~\cite{Plateau85}
of two $\rho$-reversible stochastic automata is still $\rho$-reversible and, in case of ergodicity, under
certain conditions the steady-state probability of any compound state is the product of the steady-state
probabilities of the two constituent states.

Our time reversibility result for RMPC can be rephrased in the setting of~\cite{MarinR15} in terms of
$\rho$-reversibility with $\rho$ being the identity function over states. As a consequence, the following
two results stem from Theorems~\ref{thm:time_rev_1} and~\ref{thm:time_rev_2} of the present paper and,
respectively, Theorems~$2$ and~$3$ of~\cite{MarinR15}.

	\begin{cor}[\emph{time reversibility closure}]\label{cor:closure}
Let $R_{1}, R_{2} \in \procs$ and $L \subseteq \Act$. Under the assumptions of Theorem~\ref{thm:time_rev_1}
or Theorem~\ref{thm:time_rev_2}, $\calm \lsp R_{1} \coop{L} R_{2} \rsp$ is time reversible too.
\fullbox

	\end{cor}

	\begin{cor}[\emph{product form}]\label{cor:product_form}
Let $R_{1}, R_{2} \in \procs$ and $L \subseteq \Act$. Under the assumptions of Theorem~\ref{thm:time_rev_1}
or Theorem~\ref{thm:time_rev_2}, if the set of states $\cals$ of $\calm \lsp R_{1} \coop{L} R_{2} \rsp$ is
equal to $\cals_{R_{1}} \times \cals_{R_{2}}$ where $\cals_{R_{k}}$ is the set of states of $\calm \lsp
R_{k} \rsp$ for $k \in \{ 1, 2 \}$, then $\pi(s_{1}, s_{2}) = \pi_{R_{1}}(s_{1}) \cdot \pi_{R_{2}}(s_{2})$
\linebreak for all $(s_{1}, s_{2}) \in \cals_{R_{1}} \times \cals_{R_{2}}$.
\fullbox

	\end{cor}

The product-form result above avoids the calculation of the global balance equations over $\calm \lsp R_{1}
\coop{L} R_{2} \rsp$, as $\pi(s_{1}, s_{2})$ can simply be obtained by multiplying $\pi_{R_{1}}(s_{1})$ with
$\pi_{R_{2}}(s_{2})$. However, the condition $\cals = \cals_{R_{1}} \times \cals_{R_{2}}$ requires to check
that every state in the full Cartesian product is reachable from $R_{1} \coop{L} R_{2}$. This means that no
compound state is such that its constituent states enable some action but none of these enabled actions can
be executed due to the constraints imposed by the synchronization set $L$. The condition $\cals =
\cals_{R_{1}} \times \cals_{R_{2}}$ implies that $\calm \lsp R_{1} \coop{L} R_{2} \rsp$ is ergodic over the
full Cartesian product of the two original state spaces, which is the condition used in~\cite{MarinR15}.
Although implicit in the statement of the corollary, the time reversibility of $\calm \lsp R_{1} \coop{L}
R_{2} \rsp$ is essential for the product-form result.

We conclude by observing that an important consequence of time reversibility for processes without parallel
composition, i.e., the result of Theorem~\ref{thm:time_rev_2}, and time reversibility closure with respect
to parallel composition, i.e., the result of Corollary~\ref{cor:closure}, is the fact that time
reversibility extends to the entire sublanguage $\procs'$ of RMPC in which parallel composition cannot occur
within the scope of action prefix or choice. This is quite useful because systems are typically modeled as
the parallel composition of a number of sequential processes, i.e., processes in which only action prefix
and choice can occur.

	\begin{cor}[\emph{time reversibility 3}]\label{thm:time_rev_3}
Let $R \in \procs'$. Then $\calm \lsp R \rsp$ is time reversible.
\fullbox

	\end{cor}

%
%
\section{Forward and Backward Markovian Bisimilarity}
\label{sec:bisim}
%
%

In this section, we equip RMPC with a notion of equivalence capable of identifying syntactically different
processes that expose the same observable behavior in terms of executable actions and their rates. A well
known behavioral equivalence for forward-only stochastic process calculi is Markovian
bisimilarity~\cite{Hillston96}. It equates systems that stepwise mimic each other's functional and
performance behavior and enjoys nice properties in terms of compositional reasoning as well as equational
and logical characterizations~\cite{Ber07}. We show below how it can be adapted to our reversible setting.

In the following, we work at the state space level, rather than at the linguistic level, because we address
different variants of Markovian bisimilarity that are defined not only over states, with which process terms
are naturally associated, but also over computations, which are more easily expressed through states and
transitions. We thus consider an action-labeled CTMC $(\cals, \Act \times \Rate, \reduction{\phantom{aa}})$,
where $\Act$ is a countable set of actions while $\Rate = \realns_{> 0}$ is a set of rates, which in the
case of RMPC is obtained by collecting transitions into bundles as formalized in
Definition~\ref{def:underl_ctmc}. Moreover, we use symbols $\lmp$ and $\rmp$ as multiset parentheses.

	\begin{defi}[\emph{Markovian bisimilarity}]\label{def:mb}
Two states $s_1, s_2 \in \cals$ are Markovian bisimilar, written $s_1 \mb s_2$, iff there exists a Markovian
bisimulation $\bisim$ such that $(s_1, s_2) \in \bisim$. \linebreak An equivalence relation $\bisim$ over
the set of states~$\cals$ is a Markovian bisimulation iff, whenever $(s_1, s_2) \in \bisim$, then for all
actions $a \in \Act$ and equivalence classes $C \in \cals / \bisim$:
\cws{0}{\rate(s_1, a, C) \: = \: \rate(s_2, a, C)}
where $\rate(s, a, C) \, = \, \sum \lmp \lambda \in \Rate \st \exists s' \in C \ldotp s \reduction{<a,
\lambda>} s' \rmp$.
\fullbox

	\end{defi}

\begin{figure}[t]

\centerline{\includegraphics{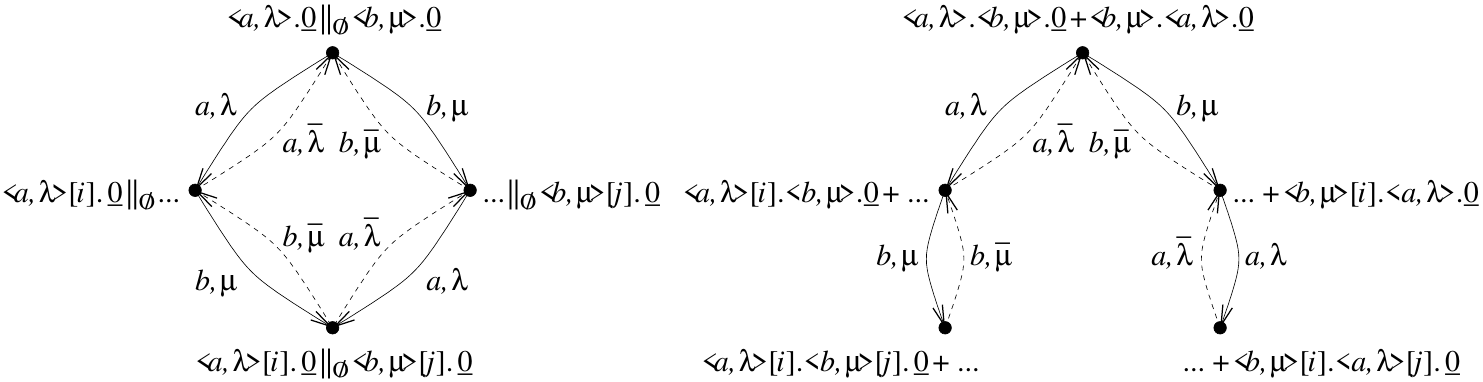}}
\caption{$\calm \lsp \lap a, \lambda \rap . \nil \coop{\emptyset} \lap b, \mu \rap . \nil \rsp$ and $\calm
\lsp \lap a, \lambda \rap . \lap b, \mu \rap . \nil + \lap b, \mu \rap . \lap a, \lambda \rap . \nil \rsp$}
\label{fig:expansion_law}

        \end{figure}

Following~\cite{ccsk}, we may adapt Markovian bisimilarity to our reversible calculus by means of two
conditions like the one in the definition above, with the former referring to forward transitions and the
latter referring to backward transitions. By so doing, we would end up with a very restrictive equivalence,
as for instance $\lap a, \lambda \rap . \nil \coop{\emptyset} \lap b, \mu \rap . \nil$ and $\lap a, \lambda
\rap . \lap b, \mu \rap . \nil + \lap b, \mu \rap . \lap a, \lambda \rap . \nil$ would be told apart. As
shown in Figure~\ref{fig:expansion_law}, in the former process from $\lap a, \lambda \rap [i] . \nil
\coop{\emptyset} \lap b, \mu \rap [j] . \nil$ both a backward $a$-transition and a backward $b$-transition
are enabled, whilst in the latter process from $\lap a, \lambda \rap [i] . \lap b, \mu \rap [j] . \nil +
\lap b, \mu \rap . \lap a, \lambda \rap . \nil$ only a backward $b$-transition is enabled as well as from
$\lap a, \lambda \rap . \lap b, \mu \rap . \nil + \lap b, \mu \rap [i] . \lap a, \lambda \rap [j] . \nil$
only a backward $a$-transition is enabled. In other words, the so-called expansion law~\cite{Milner89},
which transforms a parallel composition into a choice among all of its possible interleaved computations,
would not hold, because the two aforementioned transitions in the former process are concurrent and hence
when going backward there is no obligation to follow the path traversed in the forward direction, whereas
this is not the case with the latter process.

As recognized in~\cite{DeNicolaMV90}, in order to set up a more useful equivalence in a reversible setting,
it is necessary to enforce not only causality but also history preservation. This means that, when going
backward, a process can only move along the path representing the history that brought the process to the
current state. For example, if $\lap a, \lambda \rap . \nil \coop{\emptyset} \lap b, \mu \rap . \nil$
performs $a$ before $b$, then $a$ and $b$ can be undone in any order from $\lap a, \lambda \rap [i] . \nil
\coop{\emptyset} \lap b, \mu \rap [j] . \nil$ because there is no causality constraint between the two
actions, but history is preserved only if $b$ is undone before $a$. To accomplish this, bisimilarity has to
be defined over computations, not over states, and the notion of transition has to be suitably revised. We
start by adapting the notation of the nondeterministic setting of~\cite{DeNicolaMV90} to our stochastic
setting.

	\begin{defi}[\emph{path}]
A sequence $\xi = (s_0, \lap a_1, \lambda_1 \rap, s_1) \dots (s_{n - 1}, \lap a_n, \lambda_n \rap, s_n) \in
\: \reduction{}^*$ \linebreak is called a path of length $n$ from state $s_0$. We let $\first(\xi) = s_0$
and $\last(\xi) = s_n$; \linebreak the empty path is indicated with $\epsilon$. We denote by $\pt(s)$ the
set of paths from state $s$.
\fullbox

	\end{defi}

	\begin{defi}[\emph{run}]
A pair $\rho = (s, \xi)$ is called a run from state $s$ iff $\xi \in \pt(s)$, in which case we let
$\pt(\rho) = \xi$, $\first(\rho) = \first(\xi)$, $\last(\rho) = \last(\xi)$, with $\first(\rho) =
\last(\rho) = s$ when $\xi = \epsilon$. We denote by $\rn(s)$ the set of runs from state $s$. Given $\rho =
(s, \xi) \in \rn(s)$ and $\rho' = (s', \xi') \in \rn(s')$, their composition $\rho \rho' = (s, \xi \xi') \in
\rn(s)$ is defined iff $\last(\rho) = \first(\rho')$. We write $\rho \reduction{<a, \lambda>} \rho'$ iff
there exists $\rho'' = (s, (s, \lap a, \lambda \rap, s'))$ with $s = \last(\rho)$ such that $\rho' = \rho
\rho''$; note that $\first(\rho) = \first(\rho')$.
\fullbox

	\end{defi}

In the considered action-labeled CTMC $(\cals, \Act \times \Rate, \reduction{\phantom{aa}})$, we work with
the set $\calu$ of runs in lieu of $\cals$. Furthermore, we view the transition relation
$\reduction{\phantom{aa}}$ as being symmetric with respect to source and target states, so that every
transition can be traversed in both directions. In the setting of RMPC, this amounts to considering only the
forward transition relation thanks to Lemma~\ref{lm:loop}. Given a run $\rho$, an action $a$, and a
bisimulation equivalence class $C$, based on~\cite{DeNicolaMV90} we distinguish between the total rate of
\emph{outgoing} and \emph{incoming} run transitions, respectively, when moving between $\rho$ and $C$ via
$a$. Forward and backward Markovian bisimilarity thus relies on checking both outgoing and incoming rate
equalities.

	\begin{defi}[\emph{forward and backward Markovian bisimilarity}]\label{def:fbmb}
Two states $s_1, s_2 \in \cals$ are forward and backward Markovian bisimilar, written $s_1 \fbmb s_2$, iff
there exists a forward and backward Markovian bisimulation $\bisim$ such that $((s_1, \epsilon), (s_2,
\epsilon)) \in \bisim$. An equivalence relation $\bisim$ over the set of runs $\calu$ is a forward and
backward Markovian bisimulation iff, whenever $(\rho_1, \rho_2) \in \bisim$, then for all actions $a \in
\Act$ and equivalence classes $C \in \calu / \bisim$:
\cws{0}{\begin{array}{rcl}
\rateo(\rho_1, a, C) & \!\! = \!\! & \rateo(\rho_2, a, C) \\
\ratei(\rho_1, a, C) & \!\! = \!\! & \ratei(\rho_2, a, C) \\
\end{array}}
where:
\cws{10}{\begin{array}{rcl}
\rateo(\rho, a, C) & \!\! = \!\! & \sum \lmp \lambda \in \Rate \mid \exists \rho' \in C \ldotp \rho
\reduction{<a, \lambda>} \rho' \rmp \\
\ratei(\rho, a, C) & \!\! = \!\! & \sum \lmp \lambda \in \Rate \mid \exists \rho' \in C \ldotp \rho'
\reduction{<a, \lambda>} \rho \rmp \\
\end{array}}
\fullbox

	\end{defi}

	\begin{thm}\label{thm:fbmb}
Let $s_1, s_2 \in \cals$. Then $s_1 \fbmb s_2 \: \Longrightarrow \: s_1 \mb s_2$.

		\begin{proof}
Suppose that $s_1 \fbmb s_2$ and let $\bisim$ be a forward and backward Markovian bisimulation on $\calu$
such that $((s_1, \epsilon), (s_2, \epsilon)) \in \bisim$. We show that $\bisim' = \{ (\last(\rho_1),
\last(\rho_2)) \st (\rho_1, \rho_2) \in \bisim \}$ is a Markovian bisimulation on $\cals$, from which $s_1
\mb s_2$ will follow. \\
Consider $(\last(\rho_1), \last(\rho_2)) \in \bisim'$. By definition of $\bisim'$, we have that $(\rho_1,
\rho_2) \in \bisim$. Since $\bisim$ is a forward and backward Markovian bisimulation, for all $a \in \Act$
and $C \in \calu / \bisim$ it holds in particular that $\rateo(\rho_1, a, C) = \rateo(\rho_2, a, C)$. Since
$\rho_k \reduction{<a, \lambda>} \rho'_k$ iff $\last(\rho_k) \reduction{<a, \lambda>} \last(\rho'_k)$ for $k
\in \{ 1, 2 \}$ and -- provided that function $\last$ is lifted from runs to sets of runs -- any equivalence
class $C' \in \cals / \bisim'$ is of the form $[\last(\rho)]_{\bisim'} = \{\last(\rho') \in \cals \st
(\last(\rho), \last(\rho')) \in \bisim' \} \linebreak = \last(\{ \rho' \in \calu \st (\rho, \rho') \in
\bisim \}) = \last([\rho]_{\bisim})$, i.e., $C' = \last(C)$ for some equivalence class $C \in \calu /
\bisim$, it follows that for all $a \in \Act$ and $C' \in \cals / \bisim$ such that $C' = \last(C)$ for $C
\in \calu / \bisim$ it holds that $\rate(\last(\rho_1), a, C') = \rateo(\rho_1, a, C) = \rateo(\rho_2, a, C)
= \rate(\last(\rho_2), a, C')$.
		\end{proof}

	\end{thm}

The behavioral equivalence $\fbmb$ is strictly finer than $\mb$. Indeed, it turns out to be exceedingly
discriminating. For example, a CTMC with the only two transitions $s_{1} \reduction{<a, \lambda>} s'_{1}$
and $s_{1} \reduction{<a, \mu>} s''_{1}$ would be distinguished from a CTMC having only transition $s_{2}
\reduction{<a, \lambda + \mu>} s'_{2}$. \linebreak Observing that in terms of runs the considered
transitions are reformulated as $\rho_{1} \reduction{<a, \lambda>} \rho'_{1}$, $\rho_{1} \reduction{<a,
\mu>} \rho''_{1}$, and $\rho_{2} \reduction{<a, \lambda + \mu>} \rho'_{2}$ where $\rho_{1} = (s_{1},
\epsilon)$ and $\rho_{2} = (s_{2}, \epsilon)$, the reflexive, symmetric, and transitive closure of the
relation $\{ (\rho_{1}, \rho_{2}), (\rho'_{1}, \rho'_{2}), (\rho''_{1}, \rho'_{2}) \}$ would work well when
moving forward, as for instance $\rateo(\rho_1, a, C) = \lambda + \mu = \rateo(\rho_2, a, C)$ for $C = \{
\rho'_{1}, \rho''_{1}, \rho'_{2} \}$, whereas this would not be the case when moving backward, as for
instance $\ratei(\rho'_1, a, C) = \lambda$, $\ratei(\rho''_1, a, C) = \mu$, and $\ratei(\rho'_2, a, C) =
\lambda + \mu$ for $C = \{ \rho_{1}, \rho_{2} \}$.

This example suggests that rate-based quantitative aspects should be neglected when going backward. Indeed,
summing up rates when considering transitions departing from the same state is consistent with the fact that
the sojourn time in that state is exponentially distributed with rate given by the sum of the rates of the
outgoing transitions. Operationally, as mentioned at the end of Section~\ref{sec:integr_calculus_sem}, this
can be interpreted as if there were a race among those transitions to decide which one will be executed,
with each transition having a winning probability proportional to its rate. This race interpretation no
longer applies in the backward direction, as we sum up rates of transitions that may depart from different
states. We therefore consider a variant of $\fbmb$ that abstracts from time when going backward. It is worth
noting that, if it abstracted from time also when going forward, then we would precisely obtain the strong
back and forth bisimilarity of~\cite{DeNicolaMV90}.

	\begin{defi}[\emph{forward and time-abstract backward Markovian bisimilarity}]\label{def:ftabmb}
Two states $s_1, s_2 \in \cals$ are forward and time-abstract backward Markovian bisimilar, written $s_1
\ftabmb s_2$, iff there exists a forward and time-abstract backward Markovian bisimulation $\bisim$ such
that $((s_1, \epsilon), (s_2, \epsilon)) \in \bisim$. An equivalence relation $\bisim$ over the set of runs
$\calu$ is a forward and time-abstract backward Markovian bisimulation iff, whenever $(\rho_1, \rho_2) \in
\bisim$, then for all actions $a \in \Act$ and equivalence classes $C \in \calu / \bisim$:
\cws{0}{\begin{array}{rcl}
\rateo(\rho_1, a, C) & \!\! = \!\! & \rateo(\rho_2, a, C) \\
\transi(\rho_1, a, C) & \!\! = \!\! & \transi(\rho_2, a, C) \\
\end{array}}
where $\transi(\rho, a, C) = 1$ if there exist $\rho' \in C$ and $\lambda \in \Rate$ such that $\rho'
\reduction{<a, \lambda>} \rho$, otherwise $\transi(\rho, a, C) = 0$.
\fullbox

	\end{defi}

We conclude by showing that $\ftabmb$ -- which is defined on runs, compares for any action both outgoing
total rates and the existence of incoming transitions, and preserves history when going backward even in the
presence of concurrent transitions -- coincides with the standard $\mb$ -- which is defined on states and
compares for any action only outgoing total rates -- thus generalizing the first result
of~\cite{DeNicolaMV90} to our stochastic setting. As a consequence, when lifted to $\procs$, the former
equivalence inherits the compositionality properties and the equational and logical characterizations of the
latter~\cite{Ber07}, including in particular the expansion law as well as the stochastic variant of
idempotency according to which $\lap a, \lambda \rap . P + \lap a, \mu \rap . P$ is identified with $\lap a,
\lambda + \mu \rap . P$.

	\begin{thm}\label{thm:ftabmb}
Let $s_1, s_2 \in \cals$. Then $s_1 \ftabmb s_2 \iff s_1 \mb s_2$.

		\begin{proof}
The proof is divided into two parts:

			\begin{itemize}

\item The proof of $s_1 \ftabmb s_2 \: \Longrightarrow \: s_1 \mb s_2$ is identical to the proof of
Theorem~\ref{thm:fbmb} as only outgoing total rates are considered.

\item Suppose that $s_1 \mb s_2$. Let $\ct$ be the mapping that associates with each path $\xi$ its colored
trace, i.e., the path obtained from $\xi$ by replacing each state with its Markovian bisimulation
equivalence class:
\cws{0}{\hspace*{-0.6cm}\begin{array}{l}
\ct((s_0, \lap a_1, \lambda_1 \rap, s_1) (s_1, \lap a_2, \lambda_2 \rap, s_2) \dots (s_{n - 1}, \lap a_n,
\lambda_n \rap, s_n)) = \\
\hspace*{0.5cm} ([s_{0}]_{\mb}, \lap a_1, \lambda_1 \rap, [s_{1}]_{\mb}) ([s_{1}]_{\mb}, \lap a_2, \lambda_2
\rap, [s_{2}]_{\mb}) \dots ([s_{n-1}]_{\mb}, \lap a_n, \lambda_n \rap, [s_{n}]_{\mb}) \\
\end{array}}
Let $\bisim = \{ (\rho_1, \rho_2) \st \rho_1 \in \run(s_1), \rho_2 \in \run(s_2), \ct(\pt(\rho_1)) =
\ct(\pt(\rho_2)) \}$, which contains in particular $((s_1, \epsilon), (s_2, \epsilon))$. We show that its
reflexive, symmetric, and transitive closure $\bisim'$ is a forward and time-abstract backward Markovian
bisimulation. \\
Given $(\rho_1, \rho_2) \in \bisim$, i.e., $\rho_1 \in \run(s_1)$ and $\rho_2 \in \run(s_2)$ such that
$\ct(\pt(\rho_1)) = \ct(\pt(\rho_2))$, with $l$ being the length of $\pt(\rho_1)$ and $\pt(\rho_2)$, let us
examine the forward and backward directions respectively:

				\begin{itemize}

\item By virtue of $\ct(\pt(\rho_1)) = \ct(\pt(\rho_2))$, it holds in particular that $\last(\rho_1) \mb
\last(\rho_2)$, hence for all $a \in \Act$ and $C \in \cals / \! \mb$ we have that $\rate(\last(\rho_1), a,
C) = \rate(\last(\rho_2), a, C)$. Since $\last(\rho_k) \reduction{<a, \lambda>} \last(\rho'_k)$ iff $\rho_k
\reduction{<a, \lambda>} \rho'_k$ for $k \in \{ 1, 2 \}$, with $\rho'_k$ still belonging to $\run(s_k)$, and
any equivalence class $C' \in (\run(s_1) \cup \run(s_2)) / \bisim'$ is made of runs with paths of the same
length that traverse the same sequence of Markovian bisimulation equivalence classes of states and perform
the same sequence~of exponentially timed actions, it follows that for all $a \in \Act$ and $C' \in
(\run(s_1) \cup \run(s_2)) / \bisim'$:

					\begin{itemize}

\item if the length of all the runs of $C'$ is less than $l + 1$, then it trivially holds that
$\rateo(\rho_1, a, C') = 0 = \rateo(\rho_2, a, C')$;

\item otherwise, assuming that the states reached after $l + 1$ transitions by all of the runs of $C'$
belong to some $C \in \cals / \! \mb$, it holds that $\rateo(\rho_1, a, C') = \rate(\last(\rho_1), a, C) =
\rate(\last(\rho_2), a, C) = \rateo(\rho_2, a, C')$.

					\end{itemize}

\item Since $\ct(\pt(\rho_1)) = \ct(\pt(\rho_2))$, it cannot be the case that $\pt(\rho_1) = \epsilon$ and
$\pt(\rho_2) \neq \epsilon$, or vice versa. If $\pt(\rho_1) = \pt(\rho_2) = \epsilon$, then for all $a \in
\Act$ and $C' \in (\run(s_1) \cup \run(s_2)) / \bisim'$ it trivially holds that $\transi(\rho_1, a, C') = 0
= \transi(\rho_2, a, C')$. \\
Suppose that $\pt(\rho_1) \neq \epsilon \neq \pt(\rho_2)$, with $\rho_k = \rho'_k \rho''_k$ and $\rho''_k =
(s'_k, (s'_k, \lap a, \lambda \rap, s''_k))$, so that $s'_k \reduction{<a, \lambda>} s''_k$ is the last
transition in $\pt(\rho_k)$ and hence $\rho'_k \reduction{<a, \lambda>} \rho_k$ with $\rho'_k$ still
belonging to $\run(s_k)$, for $k \in \{ 1, 2 \}$. From $\ct(\pt(\rho_1)) = \ct(\pt(\rho_2))$, \linebreak it
follows in particular that $s'_1 \mb s'_2$. As a consequence, for all $a \in \Act$ and $C' \in (\run(s_1)
\cup \run(s_2)) / \bisim'$:

					\begin{itemize}

\item if the length of all runs of $C'$ is less than $l$, then it trivially holds that $\transi(\rho_1, a,
C') = 0 = \transi(\rho_2, a, C')$;

\item otherwise, assuming that the states reached after $l - 1$ transitions by all runs of $C'$ belong to
some $C \in \cals / \! \mb$, it holds that $\transi(\rho_1, a, C') = 1 = \transi(\rho_2, a, C')$ or
$\transi(\rho_1, a, C') = 0 = \transi(\rho_2, a, C')$ depending on whether $s'_1, s'_2 \in C$ or $s'_1, s'_2
\notin C$.
\qedhere

					\end{itemize}

				\end{itemize}

			\end{itemize}
 
		\end{proof}

	\end{thm}

%
%
\section{Examples}
\label{sec:examples}
%
%

In this section, we show how to use RMPC for modeling scenarios coming from two different application
domains: distributed systems and systems biology.

%
\subsection{Consensus Protocols}
\label{sec:consensus}
%

The two-phase commit protocol (2PC)~\cite{Mullender93} is a distributed algorithm that coordinates all the
processes participating in a distributed transaction on whether to commit or abort the transaction itself.
In the first phase, a process acting as the coordinator initiates a voting procedure to decide whether the
transaction can be committed or not. All the participants in the distributed transaction receive a request
to vote and then reply either yes for commit or no for abort depending on their local state of the
transaction. In the second phase, the coordinator decides according to the received votes. If all the
participants have agreed to commit, then the coordinator decides to commit, otherwise to abort, and notifies
its decision to the participants.

In RMPC the considered distributed system can be modeled as the parallel composition of the coordinator with
$m \ge 2$ processes participating in the transaction:
\cws{0}{\textsc{2PC} \: = \: \textsc{Coord} \coop{L} \textsc{P}_1 \coop{L} \textsc{P}_2 \coop{L} \dots
\coop{L} \textsc{P}_{m}}
where $L = \{ vt, cmt, abt \} \cup \{y_i, n_i \mid i \in I \}$, for $I = \{ 1, 2, \dots, m \}$, is the set
of actions on which the coordinator and all the participating processes have to synchronize.

The coordinator sends the request to vote at rate $\lambda$, then it waits for a yes/no answer from each
process and decides to commit only if all processes have replied yes; the commit/abort decision is broadcast
at rate $\delta$. The coordinator has a component for each possible process, with all these components being
in parallel among them with empty synchronization set (rendered as $\prod$ for brevity), and a component
counting the positive answers regardless of the order in which they arrive:
\cws{0}{\begin{array}{rcl}
\textsc{Coord} & \!\! = \!\! & \lap vt, \lambda \rap . (\textsc{Coord}' \coop{Y} \textsc{Coord}'') \\
\textsc{Coord}' & \!\! = \!\! & \prod_{i\in I} (\lap y_i, 1 \rap . \nil + \lap n_i, 1 \rap. \lap abt, \delta
\rap . \nil) \\
\textsc{Coord}'' & \!\! = \!\! & \lap y_1, 1 \rap . \lap y_2, 1 \rap . \, \dots \, . \lap y_m, 1 \rap . \lap
cmt, \delta \rap . \nil \\
\end{array}}
where $Y = \{y_i \mid i \in I \}$ and the use of rate $1$ in certain actions indicates the passivity of the
coordinator with respect to the synchronization on the corresponding activities.

After performing its part of transaction at rate $\mu_i$, the $i$-th participating process waits for a vote
request. Then at rate $\rho_i$ it answers yes with probability $p_i$, in which case it waits for the commit
or abort decision of the coordinator, or no with probability $q_i = 1 - p_i$, \linebreak in which case it
waits for the abort decision of the coordinator (let $\gamma_i = \rho_i \cdot p_i$ and $\eta_i = \rho_i
\cdot q_i$):
\cws{0}{\begin{array}{rcl}
\textsc{P}_i & \!\! = \!\! & \lap t_i, \mu_i \rap . \lap vt, 1 \rap . \textsc{P}'_i \\
\textsc{P}'_i & \!\! = \!\! & \lap y_i, \gamma_i \rap . (\lap cmt, 1 \rap . \nil + \lap abt, 1 \rap . \nil)
+ \lap n_i, \eta_i \rap . \lap abt, 1 \rap . \nil \\
\end{array}}
Even if a process chooses to commit, the coordinator may decide to abort. On the other hand, if a process
chooses to abort, it is granted that the overall transaction will be aborted.

Assuming $m = 2$ for simplicity, a standard forward computation may start with the following three
transitions and reach a term that we call $\textsc{2PC}'$:
\cws{0}{\begin{array}{rcl}
\textsc{2PC} & \!\!\! \fw{<t_1, \mu_1>[1]} \fw{<t_2, \mu_2>[2]} \fw{<vt, \lambda>[3]} \!\!\! & \lap vt,
\lambda \rap[3] . (\textsc{Coord}'
\coop{Y} \textsc{Coord}'') \coop{L} \\
& & \lap t_1, \mu_1 \rap[1] . \lap vt, 1 \rap[3] . \textsc{P}'_1 \coop{L} \\
& & \lap t_2, \mu_2 \rap[2] . \lap vt, 1 \rap[3] . \textsc{P}'_2 \\
\end{array}}
The two participating processes can now choose independently of each other. If process~$1$ votes yes while
process~$2$ votes no, so that the coordinator decides to abort the transaction, then the computation may
continue as follows and reach a term that we call
$\textsc{2PC}''$:
\cws{0}{\begin{array}{l}
\textsc{2PC}' \fw{<y_1, \gamma_1>[4]} \fw{<n_2, \eta_2>[5]} \fw{<abt, \delta>[6]} \\
\lap vt, \lambda \rap[3] . \big( \big( (\lap y_1, 1 \rap[4] . \nil + \lap n_1, 1 \rap . \lap abt, \delta
\rap . \nil) \coop{\emptyset} \\
\hspace*{2.2cm} (\lap y_2, 1 \rap . \nil + \lap n_2, 1 \rap[5] . \lap abt, \delta \rap[6] . \nil) \big)
\coop{L} \\
\hspace*{2.1cm} \lap y_1, 1 \rap[4] . \lap y_2, 1 \rap . \lap cmt, \delta \rap . \nil \big) \coop{L} \\
\lap t_1, \mu_1 \rap[1] . \lap vt, 1 \rap[3] . (\lap y_1, \gamma_1 \rap[4] . (\lap cmt, 1 \rap . \nil + \lap
abt, 1 \rap[6] . \nil) + \lap n_1, \eta_1 \rap . \lap abt, 1 \rap . \nil) \coop{L} \\
\lap t_2, \mu_2 \rap[2] . \lap vt, 1 \rap[3] . (\lap y_1, \gamma_1 \rap . (\lap cmt, 1 \rap . \nil + \lap
abt, 1 \rap . \nil) + \lap n_1, \eta_1 \rap[5] . \lap abt, 1 \rap[6] . \nil) \\
\end{array}}
where the coordinator and the two processes could only synchronize on the abort action.

At this point, the computation can only continue as a distributed rollback like the one below, which takes
place with no need to model it due to the reversibility of RMPC:
\cws{0}{\textsc{2PC}'' \bk{<abt, \past{\delta}>[6]} \bk{<n_2, \past{\eta_2}>[5]} \bk{<y_1,
\past{\gamma_1}>[4]} \bk{<vt, \past{\lambda}>[3]} \bk{<t_2, \past{\mu_2}>[2]} \bk{<t_1, \past{\mu_1}>[1]}
\textsc{2PC}}
Note that this is not the only possibility, as there are other backward computations causally equivalent to
the one above obtained by exchanging the order in which actions $y_1$ and $n_2$ \linebreak on the one hand
and actions $t_1$ and $t_2$ on the other hand are undone.

%
\subsection{Protein Interaction Networks}
\label{sec:biology}
%

Consider a prototypical model from systems biology where a molecule $A$ has multiple binding sites to which
a molecule $B$ can bind reversibly~\cite{ConzelmannFG08}. Since the number of reactions grows exponentially
with the number of binding sites, as in~\cite{SquillaceTTV22} we only examine the case of two binding sites.
We denote by $A_{1,0}$ (resp., $A_{0,1}$) a molecule $A$ having a binding with a single molecule $B$ on the
first (resp., second) site and by $A_{1,1}$ a molecule $A$ in the case that the binding is with two
molecules $B$, one on each site. In the description of the reaction network of such molecules, the left
column below indicates the binding (i.e., forward) rules while the right column below indicates the
unbinding (i.e., backward) rules where, according to the standard biochemical notation, $A + B$ stands for
the presence of at least one molecule $A$ and at least one molecule $B$ as well as $\kappa_{b_i}$ and
$\kappa_{u_i}$ represent the rates of the corresponding binding and unbinding rules:
\cws{4}{\begin{array}{rclcrcl}
A + B & \!\!\! \xrightarrow{\kappa_{b_1}} \!\!\! & A_{1,0} & \qquad &
A_{1,0} & \!\!\! \xrightarrow{\kappa_{u_1}} \!\!\! & A + B \\
A + B & \!\!\! \xrightarrow{\kappa_{b_2}} \!\!\! & A_{0,1} & \qquad &
A_{0,1} & \!\!\! \xrightarrow{\kappa_{u_2}} \!\!\! & A + B \\
A_{0,1} + B & \!\!\! \xrightarrow{\kappa_{b_1}} \!\!\! & A_{1,1} & \qquad &
A_{1,1} & \!\!\! \xrightarrow{\kappa_{u_1}} \!\!\! & A_{0,1} + B \\
A_{1,0} + B & \!\!\! \xrightarrow{\kappa_{b_2}} \!\!\! & A_{1,1} & \qquad &
A_{1,1} & \!\!\! \xrightarrow{\kappa_{u_2}} \!\!\! & A_{1,0} + B \\
\end{array}}

In RMPC we can model the same network as follows:
\cws{0}{\begin{array}{rcl}
\textsc{Net} & \!\! = \!\! & A \coop{L} (B \coop{\emptyset} B) \\
A & \!\! = \!\! & \lap b_1, 1 \rap . \nil \coop{\emptyset} \lap b_2, 1 \rap . \nil \\
B & \!\! = \!\! & \lap b_1, \kappa_{b_1} \rap . \nil + \lap b_2, \kappa_{b_2} \rap . \nil \\
\end{array}}
where $L = \{ b_1, b_2 \}$. Note that there is no need to model the unbinding rules thanks to the
reversibility of RMPC. So for instance the forward computation:
\cws{0}{\begin{array}{rcl}
\textsc{Net} & \!\! \fw{<b_1, \kappa_{b_1}>[1]} \!\! & (\lap b_1, 1 \rap[1] . \nil \coop{\emptyset} \lap
b_2, 1 \rap . \nil) \coop{L} \\
& & \big( (\lap b_1, \kappa_{b_1} \rap[1] . \nil + \lap b_2, \kappa_{b_2} \rap . \nil) \coop{\emptyset}
(\lap b_1, \kappa_{b_1} \rap . \nil + \lap b_2, \kappa_{b_2} \rap . \nil) \big) \\
& \!\! \fw{<b_2, \kappa_{b_2}>[2]} \!\! & (\lap b_1, 1 \rap[1] . \nil \coop{\emptyset} \lap b_2, 1 \rap[2] .
\nil) \coop{L} \\
& & \big( (\lap b_1, \kappa_{b_1} \rap[1] . \nil + \lap b_2, \kappa_{b_2} \rap . \nil) \coop{\emptyset}
(\lap b_1, \kappa_{b_1} \rap . \nil + \lap b_2, \kappa_{b_2} \rap[2] . \nil) \big) \\
\end{array}}
has the following two backward counterparts as the two unbindings can happen in any order from the reached
term that we call $\textsc{Net}'$:
\cws{0}{\begin{array}{c}
\textsc{Net}' \: \bk{<b_2, \past{\kappa_{b_2}}>[2]} \bk{<b_1, \past{\kappa_{b_1}}>[1]} \: \textsc{Net} \\
\textsc{Net}' \: \bk{<b_1, \past{\kappa_{b_1}}>[1]} \bk{<b_2, \past{\kappa_{b_2}}>[2]} \: \textsc{Net} \\
\end{array}}

%
%
\section{Conclusions}
\label{sec:concl}
%
%

After Landauer~\cite{Landauer61} and Bennett~\cite{Bennett73}, reversible computing has attracted a growing
attention mainly for the possibility of building energy efficient circuits. Nowadays, the interest has
spread into many application areas and hence reversible computing requires a deep investigation of its
theoretical foundations in computer science. There exist different interpretations of
reversibility in the literature. In this paper, we have addressed our research quest towards bridging causal
reversibility~\cite{rccs,DanosK05} -- developed in concurrency theory -- and time
reversibility~\cite{Kelly79} -- originated for the efficient analysis of stochastic processes.

We have accomplished this in the setting of process algebra because it constitutes a common ground for
concurrency theory and probability theory~\cite{LarsenS91,Hillston96}. Specifically, we have introduced the
stochastic process calculus RMPC, whose syntax and semantics follow the approach of~\cite{ccsk}, with the
aim of paving the way to concurrent system models that are causally reversible and time reversible \emph{by
construction}. Causal reversibility has been proved by exploiting the technique of~\cite{LanesePU20} after
importing in the setting of~\cite{ccsk} some notions coming from~\cite{rccs}. Time reversibility has been
shown under the constraint that every backward rate is equal to the corresponding forward rate -- regardless
of the syntactical structure of processes -- or that parallel composition cannot occur within the scope of
action prefix or choice -- regardless of the values of backward rates -- and has allowed us to inherit
from~\cite{MarinR15} a product form result that enables the efficient calculation of performance measures.
Finally, as far as Markovian bisimilarity~\cite{Hillston96} is concerned, we have developed two forward and
backward variants inspired by~\cite{DeNicolaMV90}, with the former preserving the expansion law and the
latter preserving a stochastic variant of the idempotency law too.

There are several lines of research that we plan to undergo, ranging from the application of our results to
case studies modeled with RMPC to the development of further theoretical results possibly admitting the
presence of irreversible actions. In particular, we would like to investigate further conditions under which
time reversibility is achieved. Our conjecture is that time reversibility should hold for the entire set of
RMPC reachable processes regardless of their backward rates and syntactical structure. This would imply that
the methodology of~\cite{ccsk} for reversing process calculi -- which we have followed in this paper -- is
fully robust not only with respect to causal reversibility, but also with respect to time reversibility.

Moreover, we wish to find a suitable way of adding recursion to the syntax of RMPC. From the point of view
of the ergodicity of the underlying CTMC, the absence of recursion is not a problem because every forward
transition has the corresponding backward transition by construction. However, there might be situations in
which recursion is necessary to appropriately describe the behavior of a system. Because of the use of
communication keys, a simple process of the form $P \triangleq \lap a, \lambda \rap . P$, whose standard
labeled transition system features a single state with a self-looping transition, produces a sequence of
infinitely many distinct states even if we resort to transition bundles. Our claim is that the specific
cooperation operator that we have considered may require a mechanism lighter than communication keys to keep
track of past actions, which may avoid the generation of an infinite state space in the presence of
recursion.

\section*{Acknowledgment}
We would like to thank Andrea Marin and Sabina Rossi for the valuable discussions on time reversibility and
the anonymous referees for their constructive comments and suggestions. This research has been supported by
the Italian MUR PRIN 2020 project \emph{NiRvAna -- Noninterference and Reversibility Analysis in Private
Blockchains}, the Italian INdAM GNCS 2022 project \emph{Propriet\`a Qualitative e Quantitative di Sistemi
Reversibili}, and the French ANR 2018 project \emph{DCore -- Causal Debugging for Concurrent Systems}.

\bibliographystyle{alphaurl}
\bibliography{biblio}

\newcommand{\etalchar}[1]{$^{#1}$}
\begin{thebibliography}{LNPV18b}

\bibitem[BAP{\etalchar{+}}12]{BerutAPCDL12}
A.~B{\'e}rut, A.~Arakelyan, A.~Petrosyan, S.~Ciliberto, R.~Dillenschneider, and
  E.~Lutz.
\newblock Experimental verification of {L}andauer's principle linking
  information and thermodynamics.
\newblock {\em Nature}, 483:187--189, 2012.

\bibitem[Ben73]{Bennett73}
C.H. Bennett.
\newblock Logical reversibility of computations.
\newblock {\em IBM Journal of Research and Development}, 17:525--532, 1973.

\bibitem[Ben03]{Bennett03}
C.H. Bennett.
\newblock Notes on {L}andauer's principle, reversible computation, and
  {M}axwell's demon.
\newblock {\em Studies in History and Philosophy of Science Part B: Studies in
  History and Philosophy of Modern Physics}, 34:501--510, 2003.

\bibitem[Ber07]{Ber07}
M.~Bernardo.
\newblock A survey of {M}arkovian behavioral equivalences.
\newblock In {\em Formal Methods for Performance Evaluation}, volume 4486 of
  {\em LNCS}, pages 180--219. Springer, 2007.

\bibitem[BKMP18]{BarylskaKMP18}
K.~Barylska, M.~Koutny, L.~Mikulski, and M.~Piatkowski.
\newblock Reversible computation vs.\ reversibility in {P}etri nets.
\newblock {\em Science of Computer Programming}, 151:48--60, 2018.

\bibitem[BM20]{BernardoM20}
M.~Bernardo and C.A. Mezzina.
\newblock Towards bridging time and causal reversibility.
\newblock In {\em Proc.\ of {FORTE~2020}}, volume 12136 of {\em LNCS}, pages
  22--38. Springer, 2020.

\bibitem[BV93]{BaetenV93}
J.C.M. Baeten and C.~Verhoef.
\newblock A congruence theorem for structured operational semantics with
  predicates.
\newblock In {\em Proc.\ of {CONCUR~1993}}, volume 715 of {\em LNCS}, pages
  477--492. Springer, 1993.

\bibitem[CFG08]{ConzelmannFG08}
H.~Conzelmann, D.~Fey, and E.D. Gilles.
\newblock Exact model reduction of combinatorial reaction networks.
\newblock {\em BMC Systems Biology}, 2(78):1--25, 2008.

\bibitem[CKV13]{CristescuKV13}
I.~Cristescu, J.~Krivine, and D.~Varacca.
\newblock A compositional semantics for the reversible $\pi$-calculus.
\newblock In {\em Proc.\ of {LICS~2013}}, pages 388--397. IEEE-CS Press, 2013.

\bibitem[DK04]{rccs}
V.~Danos and J.~Krivine.
\newblock Reversible communicating systems.
\newblock In {\em Proc.\ of {CONCUR~2004}}, volume 3170 of {\em LNCS}, pages
  292--307. Springer, 2004.

\bibitem[DK05]{DanosK05}
V.~Danos and J.~Krivine.
\newblock Transactions in {RCCS}.
\newblock In {\em Proc.\ of {CONCUR~2005}}, volume 3653 of {\em LNCS}, pages
  398--412. Springer, 2005.

\bibitem[dKH10]{deVriesKH10}
E.~{de Vries}, V.~Koutavas, and M.~Hennessy.
\newblock Communicating transactions.
\newblock In {\em Proc.\ of {CONCUR~2010}}, volume 6269 of {\em LNCS}, pages
  569--583. Springer, 2010.

\bibitem[DMV90]{DeNicolaMV90}
R.~{De Nicola}, U.~Montanari, and F.W. Vaandrager.
\newblock Back and forth bisimulations.
\newblock In {\em Proc.\ of {CONCUR~1990}}, volume 458 of {\em LNCS}, pages
  152--165. Springer, 1990.

\bibitem[Fra18]{Frank18}
M.P. Frank.
\newblock Physical foundations of {L}andauer's principle.
\newblock In {\em Proc.\ of {RC~2018}}, volume 11106 of {\em LNCS}, pages
  3--33. Springer, 2018.

\bibitem[GLM14]{GiachinoLM14}
E.~Giachino, I.~Lanese, and C.A. Mezzina.
\newblock Causal-consistent reversible debugging.
\newblock In {\em Proc.\ of {FASE~2014}}, volume 8411 of {\em LNCS}, pages
  370--384. Springer, 2014.

\bibitem[Har03]{Harrison03}
P.G. Harrison.
\newblock Turning back time in {M}arkovian process algebra.
\newblock {\em Theoretical Computer Science}, 290:1947--1986, 2003.

\bibitem[HGCH21]{HaySchmidtGCH21}
L.~Hay{-}Schmidt, R.~Gl{\"{u}}ck, M.H. Cservenka, and T.~Haulund.
\newblock Towards a unified language architecture for reversible
  object-oriented programming.
\newblock In {\em Proc.\ of {RC~2021}}, volume 12805 of {\em LNCS}, pages
  96--106. Springer, 2021.

\bibitem[Hil94]{Hillston94}
J.~Hillston.
\newblock The nature of synchronisation.
\newblock In {\em Proc.\ of {PAPM~1994}}, pages 51--70. University of Erlangen,
  Technical Report~27-4, 1994.

\bibitem[Hil96]{Hillston96}
J.~Hillston.
\newblock {\em A Compositional Approach to Performance Modelling}.
\newblock Cambridge University Press, 1996.

\bibitem[Hoa85]{Hoare85}
C.A.R. Hoare.
\newblock {\em Communicating Sequential Processes}.
\newblock Prentice Hall, 1985.

\bibitem[Jac63]{Jackson63}
J.-R. Jackson.
\newblock Jobshop-like queueing systems.
\newblock {\em Management Science}, 10:131--142, 1963.

\bibitem[Kel79]{Kelly79}
F.P. Kelly.
\newblock {\em Reversibility and Stochastic Networks}.
\newblock John Wiley \& Sons, 1979.

\bibitem[Kle75]{Kleinrock75}
L.~Kleinrock.
\newblock {\em Queueing Systems}.
\newblock John Wiley \& Sons, 1975.

\bibitem[KS60]{KemenyS60}
J.G. Kemeny and J.L. Snell.
\newblock {\em Finite {M}arkov Chains}.
\newblock Van Nostrand, 1960.

\bibitem[Lan61]{Landauer61}
R.~Landauer.
\newblock Irreversibility and heat generated in the computing process.
\newblock {\em IBM Journal of Research and Development}, 5:183--191, 1961.

\bibitem[{Lee}86]{Leeman86}
G.B. {Leeman Jr.}
\newblock A formal approach to undo operations in programming languages.
\newblock {\em {ACM} Transactions on Programming Languages and Systems},
  8:50--87, 1986.

\bibitem[LES18]{LaursenES18}
J.S. Laursen, L.{-}P. Ellekilde, and U.P. Schultz.
\newblock Modelling reversible execution of robotic assembly.
\newblock {\em Robotica}, 36:625--654, 2018.

\bibitem[L{\'{e}}v76]{Levy76}
J.-J. L{\'{e}}vy.
\newblock An algebraic interpretation of the $\lambda\beta${K}-calculus; and an
  application of a labelled $\lambda$-calculus.
\newblock {\em Theoretical Computer Science}, 2:97--114, 1976.

\bibitem[LLM{\etalchar{+}}13]{LaneseLMSS13}
I.~Lanese, M.~Lienhardt, C.A. Mezzina, A.~Schmitt, and J.-B. Stefani.
\newblock Concurrent flexible reversibility.
\newblock In {\em Proc.\ of {ESOP~2013}}, volume 7792 of {\em LNCS}, pages
  370--390. Springer, 2013.

\bibitem[LLMS12]{LienhardtLMS12}
M.~Lienhardt, I.~Lanese, C.A. Mezzina, and J.-B. Stefani.
\newblock A reversible abstract machine and its space overhead.
\newblock In {\em Proc.\ of {FMOODS/FORTE~2012}}, volume 7273 of {\em LNCS},
  pages 1--17. Springer, 2012.

\bibitem[LM20]{LaneseM20}
I.~Lanese and D.~Medi{\'c}.
\newblock A general approach to derive uncontrolled reversible semantics.
\newblock In {\em Proc.\ of {CONCUR~2020}}, volume 171 of {\em LIPIcs}, pages
  33:1--33:24. Schloss Dagstuhl - Leibniz-Zentrum f{\"{u}}r Informatik, 2020.

\bibitem[LMM21]{LaneseMM21}
I.~Lanese, D.~Medi{\'c}, and C.A. Mezzina.
\newblock Static versus dynamic reversibility in {CCS}.
\newblock {\em Acta Informatica}, 58:1--34, 2021.

\bibitem[LMS10]{LaneseMS10}
I.~Lanese, C.A. Mezzina, and J.-B. Stefani.
\newblock Reversing higher-order $\pi$.
\newblock In {\em Proc.\ of {CONCUR~2010}}, volume 6269 of {\em LNCS}, pages
  478--493. Springer, 2010.

\bibitem[LNPV18a]{LaneseNPV18a}
I.~Lanese, N.~Nishida, A.~Palacios, and G.~Vidal.
\newblock Cau{DE}r: {A} causal-consistent reversible debugger for {E}rlang.
\newblock In {\em Proc.\ of {FLOPS~2018}}, volume 10818 of {\em LNCS}, pages
  247--263. Springer, 2018.

\bibitem[LNPV18b]{LaneseNPV18b}
I.~Lanese, N.~Nishida, A.~Palacios, and G.~Vidal.
\newblock A theory of reversibility for {E}rlang.
\newblock {\em Journal of Logical and Algebraic Methods in Programming},
  100:71--97, 2018.

\bibitem[LPU20]{LanesePU20}
I.~Lanese, I.C.C. Phillips, and I.~Ulidowski.
\newblock An axiomatic approach to reversible computation.
\newblock In {\em Proc.\ of {FOSSACS~2020}}, volume 12077, pages 442--461.
  Springer, 2020.

\bibitem[LS91]{LarsenS91}
K.G. Larsen and A.~Skou.
\newblock Bisimulation through probabilistic testing.
\newblock {\em Information and Computation}, 94:1--28, 1991.

\bibitem[Mil89]{Milner89}
R.~Milner.
\newblock {\em Communication and Concurrency}.
\newblock Prentice Hall, 1989.

\bibitem[MR15]{MarinR15}
A.~Marin and S.~Rossi.
\newblock Quantitative analysis of concurrent reversible computations.
\newblock In {\em Proc.\ of {FORMATS~2015}}, volume 9268 of {\em LNCS}, pages
  206--221. Springer, 2015.

\bibitem[Mul93]{Mullender93}
S.~Mullender.
\newblock {\em Distributed Systems}.
\newblock Addison-Wesley, 1993.

\bibitem[Pin17]{Pinna17}
G.M. Pinna.
\newblock Reversing steps in membrane systems computations.
\newblock In {\em Proc.\ of {CMC~2017}}, volume 10725 of {\em LNCS}, pages
  245--261. Springer, 2017.

\bibitem[Pla85]{Plateau85}
B.~Plateau.
\newblock On the stochastic structure of parallelism and synchronization models
  for distributed algorithms.
\newblock In {\em Proc.\ of {SIGMETRICS~1985}}, pages 147--154. ACM Press,
  1985.

\bibitem[PP14]{PerumallaP14}
K.S. Perumalla and A.J. Park.
\newblock Reverse computation for rollback-based fault tolerance in large
  parallel systems - {E}valuating the potential gains and systems effects.
\newblock {\em Cluster Computing}, 17:303--313, 2014.

\bibitem[PU07]{ccsk}
I.C.C. Phillips and I.~Ulidowski.
\newblock Reversing algebraic process calculi.
\newblock {\em Journal of Logic and Algebraic Programming}, 73:70--96, 2007.

\bibitem[PUY13]{PhillipsUY12}
I.C.C. Phillips, I.~Ulidowski, and S.~Yuen.
\newblock A reversible process calculus and the modelling of the {ERK}
  signalling pathway.
\newblock In {\em Proc.\ of {RC~2012}}, volume 7581 of {\em LNCS}, pages
  218--232. Springer, 2013.

\bibitem[PV15]{PalaciosV15}
A.~Palacios and G.~Vidal.
\newblock Concolic execution in functional programming by program
  instrumentation.
\newblock In {\em Proc.\ of {LOPSTR~2015}}, volume 9527 of {\em LNCS}, pages
  277--292. Springer, 2015.

\bibitem[SOJB18]{SchordanOJB18}
M.~Schordan, T.~Oppelstrup, D.R. Jefferson, and P.D. {Barnes Jr.}
\newblock Generation of reversible {C++} code for optimistic parallel discrete
  event simulation.
\newblock {\em New Generation Computing}, 36:257--280, 2018.

\bibitem[SPP19]{SiljakPP19}
H.~Siljak, K.~Psara, and A.~Philippou.
\newblock Distributed antenna selection for massive {MIMO} using reversing
  {P}etri nets.
\newblock {\em IEEE Wireless Communication Letters}, 8:1427--1430, 2019.

\bibitem[Ste94]{Stewart94}
W.J. Stewart.
\newblock {\em Introduction to the Numerical Solution of {M}arkov Chains}.
\newblock Princeton University Press, 1994.

\bibitem[STTV22]{SquillaceTTV22}
G.~Squillace, M.~Tribastone, M.~Tschaikowski, and A.~Vandin.
\newblock An algorithm for the formal reduction of differential equations as
  over-approximations.
\newblock In {\em Proc.\ of {QEST~2022}}, volume 13479 of {\em LNCS}, pages
  173--191. Springer, 2022.

\bibitem[VS18]{VassorS18}
M.~Vassor and J.-B. Stefani.
\newblock Checkpoint/rollback vs causally-consistent reversibility.
\newblock In {\em Proc.\ of {RC~2018}}, volume 11106 of {\em LNCS}, pages
  286--303. Springer, 2018.

\bibitem[YG07]{YokoyamaG07}
T.~Yokoyama and R.~Gl{\"{u}}ck.
\newblock A reversible programming language and its invertible
  self-interpreter.
\newblock In {\em Proc.\ of {PEPM~2007}}, pages 144--153. ACM Press, 2007.

\end{thebibliography}

\end{document}